\documentclass{aa}
\usepackage[fleqn]{amsmath}
\usepackage{graphicx,natbib,txfonts,color,upgreek}
\bibpunct{(}{)}{;}{a}{}{,}

\begin{document}
\title{On the structure of the transition disk around TW\,Hya}
\author{J. Menu\inst{1}\fnmsep\inst{2}\fnmsep\thanks{PhD fellow of the Research Foundation -- Flanders (FWO)} \and R. van Boekel\inst{2} \and T. Henning\inst{2} \and C.~J.~Chandler\inst{3} \and H. Linz\inst{2} \and M. Benisty\inst{4} \and S. Lacour\inst{5} \and M. Min\inst{6} \and C.~Waelkens\inst{1} \and S.~M.~Andrews\inst{7} \and N.~Calvet\inst{8} \and J.~M.~Carpenter\inst{9} \and S.~A.~Corder\inst{10} \and A.~T.~Deller\inst{11} \and J.~S.~Greaves\inst{12} \and R.~J.~Harris\inst{7} \and A.~Isella\inst{9} \and W.~Kwon\inst{13} \and J.~Lazio\inst{14} \and J.-B.~Le~Bouquin\inst{4}\and F.~M\'enard\inst{15,4} \and L.~G.~Mundy\inst{16}  \and L.~M.~P\'erez\inst{3}\fnmsep\thanks{Jansky Fellow, National Radio Astronomy Observatory} \and L.~Ricci\inst{9} \and A.~I.~Sargent\inst{9} \and S.~Storm\inst{16} \and L.~Testi\inst{17,18} \and D.~J.~Wilner\inst{7}
}

\institute{
Instituut voor Sterrenkunde, KU Leuven, Celestijnenlaan 200D, 3001 Leuven, Belgium \\ \email{jonathan.menu@ster.kuleuven.be} 
\and
Max Planck Institut f\"ur Astronomie, K\"onigstuhl 17, 69117 Heidelberg, Germany
\and
National Radio Astronomy Observatory, P.O. Box O, Socorro, NM 87801, USA
\and
CNRS/UJF Grenoble 1, UMR 5274, Institut de Plan\'etologie et d'Astrophysique de Grenoble (IPAG), 38041 Grenoble, France
\and
LESIA, Observatoire de Paris, CNRS, UPMC, Universit\'e Paris Diderot, 5 place Jules Janssen, 92195 Meudon, France
\and
Astronomical Institute ``Anton Pannekoek'', University of Amsterdam, PO Box 94249, 1090 GE Amsterdam, The Netherlands
\and
Harvard-Smithsonian Center for Astrophysics, 60 Garden Street, Cambridge, MA 02138, USA
\and
Department of Astronomy, University of Michigan, 830 Dennison Building, 500 Church Street, Ann Arbor, MI 48109, USA
\and
Astronomy Department, California Institute of Technology, 1200 East California Boulevard, Pasadena, CA 91125, USA
\and
Joint ALMA Observatory, Av. Alonso de C\'ordova 3107, Vitacura, Santiago, Chile
\and
The Netherlands Institute for Radio Astronomy (ASTRON), 7990-AA Dwingeloo, The Netherlands
\and
School of Physics and Astronomy, University of St. Andrews, North Haugh, St Andrews KY16 9SS, UK
\and
SRON Netherlands Institute for Space Research, Landleven 12, 9747 AD Groningen, The Netherlands
\and
Jet Propulsion Laboratory, California Institute of Technology, 4800 Oak Grove Drive, Pasadena, CA 91106, USA
\and
UMI-FCA, CNRS/INSU France (UMI 3386), and Departamento de Astronom\'ia, Universidad de Chile, Santiago, Chile
\and
Department of Astronomy, University of Maryland, College Park, MD 20742, USA
\and
European Southern Observatory, Karl Schwarzschild Str. 2, 85748 Garching, Germany
\and
INAF-Osservatorio Astrofisico di Arcetri, Largo E. Fermi 5, 50125 Firenze, Italy
}

\date{Received / Accepted}

\authorrunning{J.~Menu et al.}
\titlerunning{TW\,Hya: multiwavelength interferometry of a transition disk}

\abstract
{For over a decade, the structure of the inner cavity in the transition disk of TW\,Hydrae has been a subject of debate. Modeling the disk with data obtained at different wavelengths has led to a variety of proposed disk structures. Rather than being inconsistent, the individual models might point to the different faces of physical processes going on in disks, such as dust growth and planet formation.}
{Our aim is to investigate the structure of the transition disk again and to find to what extent we can reconcile apparent model differences.}
{A large set of high-angular-resolution data was collected from near-infrared to centimeter wavelengths. We investigated the existing disk models and established a new self-consistent radiative-transfer model. A genetic fitting algorithm was used to automatize the parameter fitting, and uncertainties were investigated in a Bayesian framework.}
{Simple disk models with a vertical inner rim and a radially homogeneous dust composition from small to large grains cannot reproduce the combined data set. Two modifications are applied to this simple disk model: (1)~the inner rim is smoothed by exponentially decreasing the surface density in the inner $\sim3\,$AU, and (2)~the largest grains ($>100\,\mu$m) are concentrated towards the inner disk region. Both properties can be linked to fundamental processes that determine the evolution of protoplanetary disks: the shaping by a possible companion and the different regimes of dust-grain growth, respectively. }
{The full interferometric data set from near-infrared to centimeter wavelengths requires a revision of existing models for the TW\,Hya disk. We present a new model that incorporates the characteristic structures of previous models but deviates in two key aspects: it does not have a sharp edge at 4\,AU, and the surface density of large grains differs from that of smaller grains. This is the first successful radiative-transfer-based model for a full set of interferometric data.}

\keywords{protoplanetary disks, techniques: interferometric, stars: individual (TW\,Hya)}

\maketitle


\section{Introduction}
With the potential discovery of (proto)planetary bodies in disks around young stars \citep{2012ApJ...745....5K,2013ApJ...766L...1Q}, observations started to provide the ultimate justification for calling the disks ``protoplanetary''. The conservation of angular momentum naturally explains why the collapsing cloud around stars evolves to a much more compact rotating disk of dust and gas. In contrast, the subsequent evolution of this protoplanetary disk towards a gas-free debris disk or a genuine planetary system is a much more complicated---and less understood---process (for an overview, see, e.g.,~\citealt{2011ppcd.book..114H,2011ARA&A..49...67W}).

The current hypothesis on protoplanetary disk evolution includes a short dissipation phase, in which disk material disperses outwards. A substantial fraction of disks around young stars is indeed observed to have an inner gap or ``hole'', a population of so-called \emph{transition disks}. At least four mechanisms have been proposed for the dissipation mechanism. First, \emph{photoevaporation} might be responsible for physically clearing the inner-disk material, a process driven by energetic photons (e.g.,~\citealt{2009ApJ...705.1237G,2010MNRAS.401.1415O}). A second mechanism is the \emph{dynamical clearing} by a (sub-) stellar companion (e.g.,~\citealt{1994ApJ...421..651A}). Third, \emph{magneto-rotational instability} might take place in an ionized disk rim, and therefore drain out the inner disk (e.g.,~\citealt{2007NatPh...3..604C}). Finally, the inner gap could actually only be an opacity gap caused by \emph{grain growth}  (e.g.,~\citealt{2005A&A...434..971D}), a process diminishing the emitting/absorbing surface of the dust and therefore creating (only) an \emph{apparent} gap. Recent simulations \citep{2011MNRAS.412...13O} suggest that a combination of several processes might be necessary to explain all cases. 

The observational situation on transition-disk objects is complex. Recent mm-observations confirm low densities in mm-sized grains in the inner disk of a sample of transition disks \citep{2011ApJ...732...42A}, but modeling of these data predicts much stronger depletion in $\mu$m-sized grains than actually observed \citep{2012ApJ...750..161D}. In the inner regions, several ``standard'' transition-disk sources seem to differ significantly. Based on scattered-light images and SED-modeling, \citet{2012ApJ...752..143H} model the central cavity of HD\,169142 as being almost completely empty, with only a very compact disk and a halo-like central region consisting of pure carbonaceous material. SED-modeling of the disk of LkCa 15 leads to similar geometries, but for this model, the compact component is predominantly filled with large silicate grains \citep{2010A&A...512A..11M}. A third case is T Cha, for which \citet{2011A&A...528L...6O} model near-IR interferometry and SED data with a tiny inner disk of small silicate and carbon grains. Similarly, HD\,100546 is modeled as having a tiny inner disk, but the outer disk of this particular object is found to have a rounded rim \citep{2013A&A...557A..68M}. Seemingly, for the handful of transition disks that have been modeled in detail, there is no consistent relation found on dust composition, grain sizes and inner-disk geometry.

Since it was identified to be a transition disk, the disk around TW Hydrae (TW\,Hya) has been one of the principal objects for investigating the physical conditions in dispersing protoplanetary disks. Various studies reveal gas lines of different molecules, allowing us to constrain the physical properties of the gas in the disk (e.g.,~\citealt{2010A&A...518L.125T,2011ApJ...735...90G,2011ApJ...736...13P,2012ApJ...748....6G,2012ApJ...757..129R}). The discovery of a handful of key molecules in the disk, such as gaseous H$_2$O \citep{2011Sci...334..338H} and HD \citep{2013Natur.493..644B}, allows pioneering work to be done in the study of early solar-system analogues. The latter work also led to an estimate of the total disk mass.

Several of the mentioned works start from a certain disk geometry for constraining model quantities. The geometry in its turn is derived from the dust emission profile, the most important diagnostic for constraining structural parameters of the disk. However, as we will indicate in Sect.~\ref{sect:history}, a consensus on the disk structure is actually far from being reached. Different instruments operating at different wavelengths and resolutions probe different regions of the disk both in terms of radial distance to the central star and vertical distance above the disk midplane. In addition, they may be sensitive to very different parts of the dust (size) distribution. This can lead to large differences between models fit to different observations of the same object, even if each model adequately reproduces the data it was meant to explain. As a result, the full dust distribution in the disk, which forms the backbone for radiation balance and therefore the disk temperature structure, remains incompletely constrained. 

We have collected an extended set of new and archival high angular resolution data for reinvestigating the TW\,Hya disk. In Sect.~\ref{sect:history} we give a short overview of disk-structure models that have been published for TW\,Hya. Sect.~\ref{sect:observations} introduces the data set and comments on the data reduction, while a first qualitative analysis of the data is given in Sect.~\ref{sect:inspection}. We confront several previous disk models with the full data set in Sect.~\ref{sect:reimplementing} and discuss potential shortcomings. This opens the way to Sect.~\ref{sect:modeling}, where we present our detailed radiative-transfer analysis. In Sect.~\ref{sect:discussion}, the results are being discussed, and Sect.~\ref{sect:nearinfrared} goes in some more detail considering the analysis of the near-infrared data. Finally, Sect.~\ref{sect:conclusion} summarizes the conclusions of this work.


\section{The structure of the TW\,Hya disk}\label{sect:history}
Although there had been earlier attempts to infer \emph{indirect} spatial information on the TW\,Hya star-disk system (e.g.,~\citealt{1993A&A...268..624B}; \citealt{1997Sci...277...67K}; \citealt{1998A&A...340..135M}), Hubble Space Telescope (HST) and millimeter imaging had to be awaited for \emph{directly} studying the TW\,Hya protoplanetary disk. The first HST images were presented by \citet{2000ApJ...538..793K} and \citet{2002ApJ...566..409W}, indicating a nearly pole-on orientation and significant radial brightness variations in an otherwise symmetric disk. New HST imaging by \citet{2005ApJ...622.1171R} confirmed the disk orientation and reported an outer radius of at least 280\,AU in scattered light. This work also reported a slight azimuthal asymmetry, possibly due to a warp. Also \citet{2004A&A...415..671A} imaged the disk, this time using ground-based polarimetry. Meanwhile, \citet{2000ApJ...534L.101W} had shown the disk to be resolved at 7\,mm, and \citet{2004ApJ...616L..11Q} had put quantitative constraints on the inclination ($i\sim7^\circ$) using CO emission line data.

A major leap forward in the study of the protoplanetary disk came with the work of \citet{2002ApJ...568.1008C}. On the basis of detailed SED modeling, these authors derived a first complete set of disk parameters, including the disk's mass and dust properties. Perhaps the most interesting result of this work is the introduction of a dust-depleted inner hole of radius $4\,$AU in the disk model.\footnote{An inner disk edge that is larger than the dust sublimation radius was already used in a disk model in \citet{2001ApJ...552L.151T}, although not interpreted as a departure from a classical disk structure.} One of the speculations of the authors for this gap was the presence of a growing exoplanet: ``TW\,Hya may become the Rosetta stone for our understanding of the evolution and dissipation of protoplanetary disks'' \citep{2002ApJ...568.1008C}.

The new transition-disk status made TW\,Hya an attractive target for new detailed high spatial resolution observations. Near-infrared interferometric observations by \citet{2006ApJ...637L.133E} are interpreted as being evidence of an inner optically-thin region in the range $0.06$ to $4\,$AU, consisting of small grains. Also \citet{2007ApJ...664..536H} find their VLA mm-interferometry to be consistent with a 4-AU gap disk. Interestingly, mid-infrared interferometric observations published by \citet{2007A&A...471..173R} point to a different gap size: these authors find a disk model with an $\sim0.7$\,AU gap radius. 

In an attempt to reconcile the apparent ambiguity in the gap sizes, \citet{2011ApJ...728...96A} performed a combined analysis of the three above interferometric data sets. Using a simple disk model, these authors reconfirmed the 4-AU radius gap, but an additional optically thick ring with a 0.5-AU radius needed to be introduced in the model. Thus, the different characteristic sizes inferred in the previous studies would not necessarily be in contradiction with each other. The most recent study for the structure of the inner-disk region is based on mid-infrared speckle interferometry, and is presented by \citet{2012ApJ...750..119A}. Again, the large gap size is confirmed, but these authors argue the presence of a hotter-than-equilibrium source at $\sim3.5\,$AU for explaining their data and the \citet{2007A&A...471..173R} data. A different aspect is that these authors reveal variability in the mid-infrared visibility curves.

Finally, recent work has shown the TW\,Hya disk also to be more complex on larger scales. Using sub-mm observations, \citet{2012ApJ...744..162A} show that the distribution of the bulk of the large-grain circumstellar dust has a sharp outer cutoff at $60\,$AU, whereas the distribution in CO gas continues to over 200\,AU. Given the constraints on scattered light from HST images, these different scales might point to a spatial separation between large and small dust grains, the latter coupled to the gas. \citet{2013ApJ...771...45D}, on the other hand, detect a gap in the disk at $\sim80\,$AU using scattered-light observations. The mechanisms causing the sharp changes in (dust) surface density at these large disk scales are yet to be determined.

\section{Observations and data reduction}\label{sect:observations}
In this work, we will make use of both new and published (or archival) data sets obtained with different instruments, at different wavelengths. In summary, this is near-infrared interferometry (VLTI/PIONIER), near-infrared sparse aperture masking (VLT/NaCo), mid-infrared interferometry (VLTI/MIDI), sub-mm interferometry (SMA), mm/cm-interferometry (VLA), and spectral energy distribution data. We give an overview of the different data sets below, and order the high angular resolution data according to increasing wavelength.

\subsection{Near-infrared interferometry and sparse-aperture masking}
The PIONIER observations of TW\,Hya were performed at the Paranal Observatory of the European Southern Observatory (ESO) on February 2, 2011. The weather conditions were clear with average seeing. The four 1.8-m Auxiliary Telescopes (ATs) of the Very Large Telescope Interferometer (VLTI; \citealt{2010SPIE.7734E...3H}) were used in extended array configuration and located at the following stations: A0-K0-G1-I1.  The use of four telescopes simultaneously allows for measurements of six simultaneous baselines, leading to three independent closure phase measurements as well. 

The observation strategy was designed to intertwine the science target TW\,Hya between two different interferometric calibrators, here HIP\,53487 and HIP\,54547. The observation sequence (five blocks) was Calibrator 1 -- Science Target -- Calibrator 2 -- Science Target -- Calibrator 1. Each block, either science or calibrator,  was composed of five exposures, each of which composed of 100 scans.  

The data were reduced and calibrated by running the \texttt{pndrs} package  on the data and calibration files. Selection was applied to reject the interferograms with signal-to-noise ratio below a threshold. Typical errors on individual measurements are $\sim3\,\%$ for the visibilities and $\sim3$ degrees on the closure phases. For a detailed description of the instrument and the data reduction pipeline, we refer to \citet{2011A&A...535A..67L}.

The sparse aperture masking (SAM) data were obtained with the NaCo-instrument on Unit Telescope 4 (UT4) of the Very Large Telescope (VLT). The observations were performed in March 2009 using the L prime filter (L$'$, 3.8 $\mu$m) and the infrared Adaptive-Optics (AO) wavefront sensor, and the seeing conditions were good ($\sim 0.7\,$arcsec). 

The total observation time was 2.5 hours with 80\,min.~of effective integration. It consisted of five Science target -- Calibrator pairs of eight datacubes each, using three different 2MASS calibrators of a magnitude within 0.5\,mag of TW\,Hya (2MASS J11023861-3433309, 2MASS 11000281-3506367 and 2MASS J11001754-3529224). The second calibrator was found to be a binary system with $\Delta$L$'\sim 3$\,mag and was therefore discarded. 

From this dataset, the closure phases were previously computed and analyzed in \citet{2012ApJ...744..120E}. The closure phases were found to be consistent with 0, pointing to a point-symmetric disk geometry. In the current paper, we focus on the visibilities to constrain the (point-symmetric) geometry of the disk. We reprocessed the data with the \texttt{SAMP} pipeline \citep{2011A&A...532A..72L}. It includes sky subtraction, bad pixel subtraction and fringe fitting. The visibility-square measurements are then normalized by the stellar flux, calibrated by the visibilities obtained on the two calibrators, and azimuthally averaged (i.e.,~over the identical baseline lengths). The latter step is justified given the (almost) pole-on orientation of the disk. Systematic errors were checked by calibrating the data of the faint calibrator with the other (2MASS J11001754-3529224 is 0.4\,mag brighter in H and K band than 2MASS J11023861-3433309). No dependencies on the brightness were found within our error bars, which means that, if the quality of the AO correction is a function of the brightness of the source, it was not noticeable for the range of brightness of our sources.

\begin{figure}
 \includegraphics[width=.5\textwidth,viewport=0 10 430 410,clip]{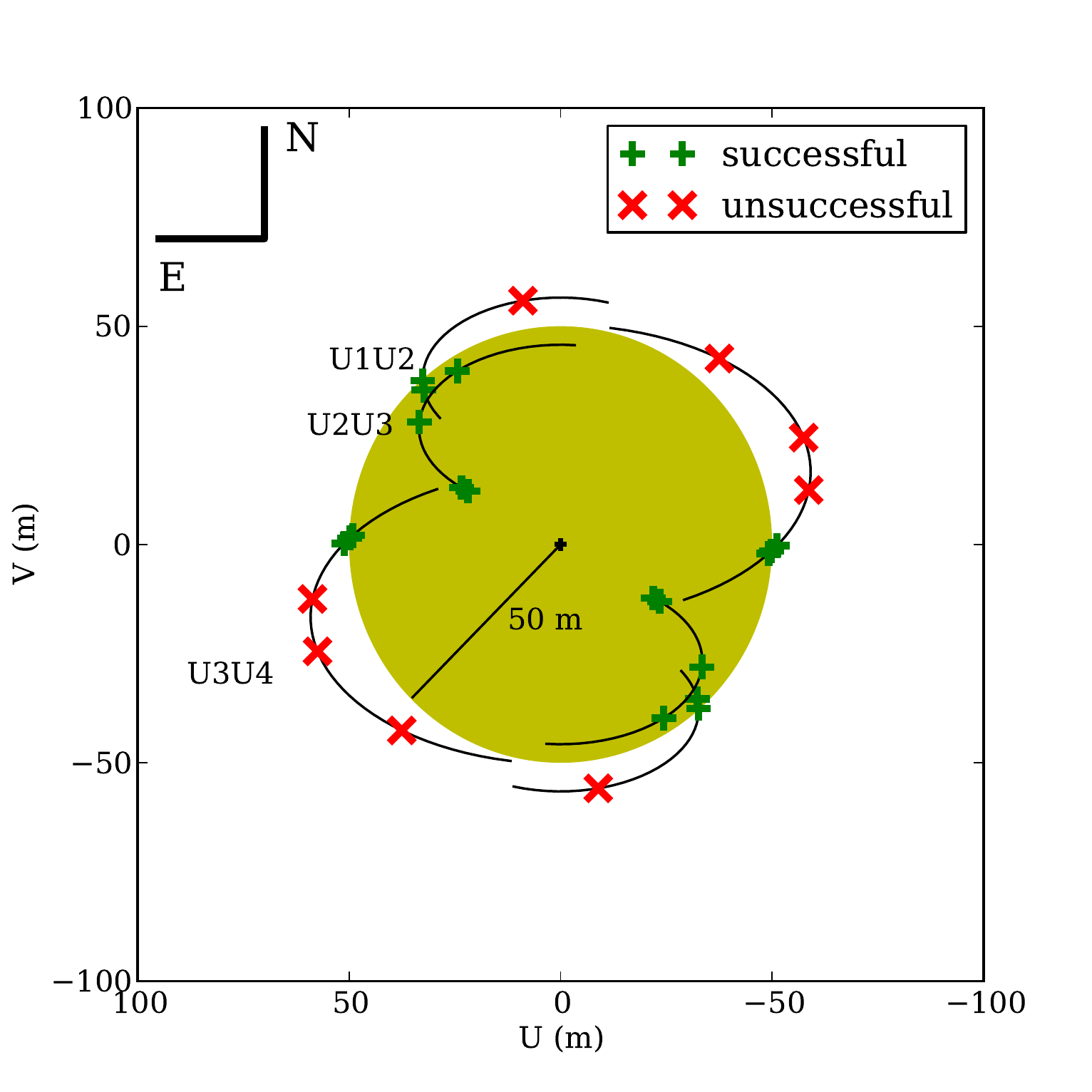}
 \caption{Uv-plot of all MIDI observations of TW\,Hya, distinguishing between the successful (green) and unsuccessful (red) fringe tracks. The curved tracks correspond to the different baselines, with the different names indicated. Clearly, the yellow area with a 50-m radius indicates the domain accessible with MIDI.}\label{fig:UV}
\end{figure}

\subsection{Mid-infrared interferometry}
TW\,Hya was observed with VLTI/MIDI during several epochs in the period 2005-2011. The projected baselines during these observations ranged from 25\,m to 70\,m, a plot of the $(u,v)$ plane is shown in Fig.~\ref{fig:UV}. With a total flux below 1\,Jy at 10\,$\mu$m, the object is at the limit of what is possible with MIDI on the Unit Telescopes. Not all attempted observations were successful, as is indicated in the figure. 

We made use of version 2.0 of the \texttt{EWS} software \citep{2004SPIE.5491..715J}, released in October 2012, for the data reduction. \texttt{EWS} is based on the coherent estimation of the visibility signal. The new version is especially designed to maximize the performance on weak targets. A standard data reduction now involves a stepwise, improved determination of the optical path differences (OPDs), allowing a better cophasing and averaging of the visibility signal \citep{2012SPIE.8445E..1GB}. 

For very faint objects, as is the case for TW\,Hya, the estimation of the OPD can still not be expected to be as accurate as for the used calibrator. This might slightly degrade the cophasing of the visibility signal, resulting in a lower calibrated flux than expected. We can correct for these correlation losses by performing a ``dilution experiment'' on a bright calibrator, as described in \citet{2012SPIE.8445E..1GB}. In short, the calibrator signal in the raw calibrator data is artificially diluted to the same level as that of the science target (i.e.,~TW\,Hya), and these artificial data are reduced again. Comparing the newly reduced data with the expected result, a correlation-loss factor can be defined for upscaling the science data. Values we found for the correlation losses were $5\,\%$ to $25\,\%$, depending on the original correlated-flux level and the observing conditions. These values agree with statistics in \citet{2012SPIE.8445E..1GB} for a range of correlation-loss experiments.

The two 2005 observations on a 50-m baseline were already published by \citet{2007A&A...471..173R}. For the reduction, the authors make use of a previous version of the data reduction software used here. The correlated fluxes found in the current paper differ slightly towards the red edge of the N-band. Given that our data reduction is based on a more recent version of the software, especially designed for faint objects, we believe that the newly reduced data are of a higher quality. In particular, the N-band ozone feature often affecting the quality of mid-infrared data in the range $9.2-10\,\mu$m seems to be ``calibrated out'', indicative of a good quality calibration.

\subsection{Sub-mm interferometry}
Since the goal of this paper is to get a coherent view of the dust distribution in the complete TW\,Hya disk, observational constraints on the large scale are as essential as constraints on the small scale. We therefore include the continuum SMA data published in \citet{2012ApJ...744..162A}, mainly probing the outer regions of the disk. We restrict ourselves to modeling the deprojected, azimuthally averaged real visibility profile (the deprojected imaginary component is effectively zero).

\begin{figure}
 \includegraphics[width=.5\textwidth]{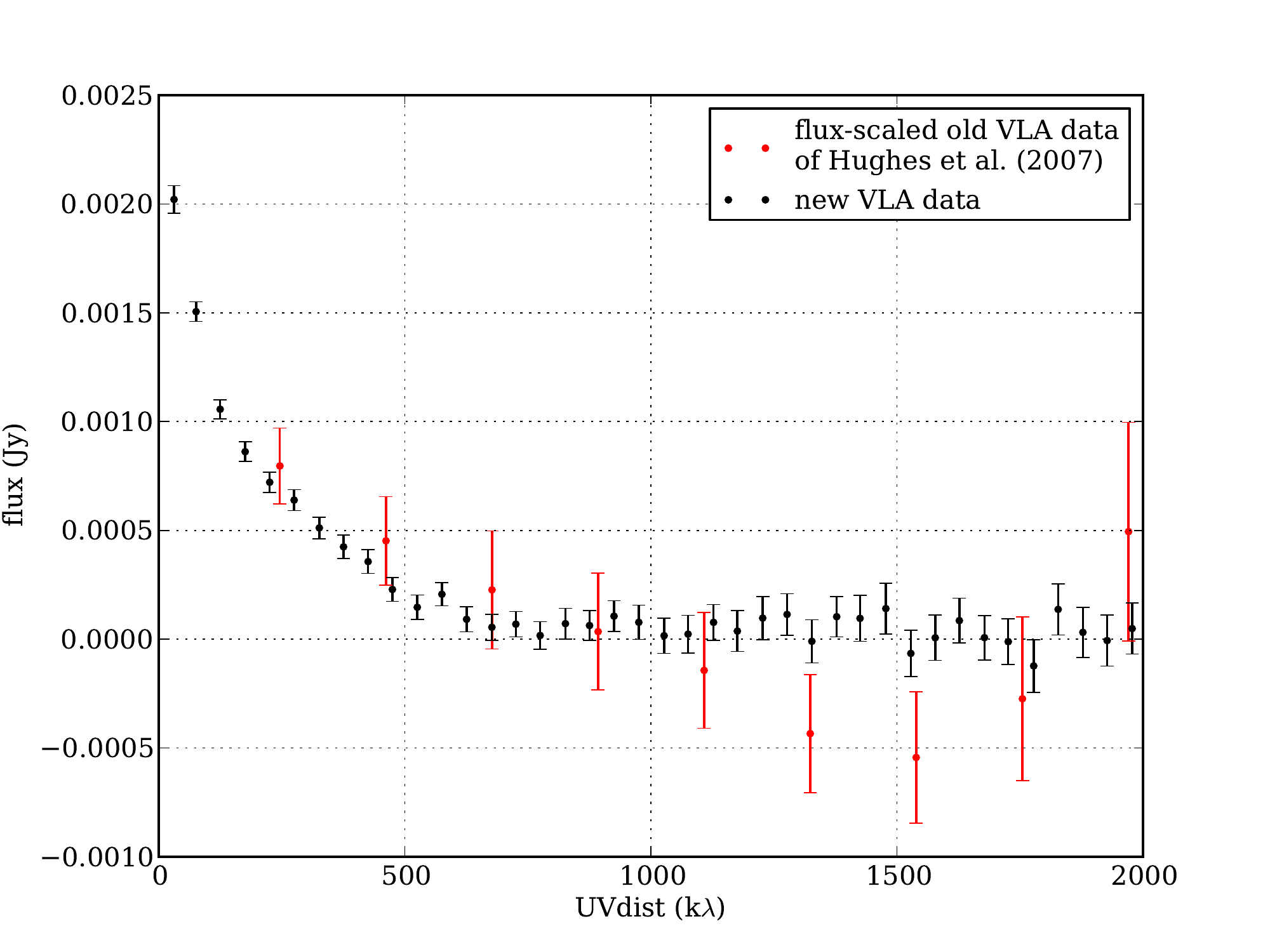}
 \caption{Comparison of the new VLA data with the previous 7-mm VLA data set of \citet{2007ApJ...664..536H}. The flux for the latter was rescaled applying the spectral index of $\alpha=2.47$ to allow the comparison with our $9.315$-mm data. Note that the bins for the previous data set are overlapping: only non-adjacent points are effectively independent (see \citealt{2007ApJ...664..536H}).}\label{fig:oldandnewVLA}
\end{figure}

\subsection{mm/cm-interferometry}\label{sect:mminterfero}
TW\,Hya was observed with the Karl G.~Jansky Very Large Array (VLA) as part of the Disks$@$EVLA program (AC982, PI C.~Chandler).  The data presented here were obtained in two 1-GHz Intermediate Frequency channels (IFs) centered at 30.5 and 37.5\,GHz (9.8 and 7.9\,mm) simultaneously, using the Ka-band receivers, in the CnB, BnA, and A configurations, between 2011 January and 2012 October (see \citealt{2011ApJ...739L...1P} for an overview of the new capabilities of the VLA).  The data were calibrated using the CASA data reduction package (e.g., \citealt{2007ASPC..376..127M}), and the VLA Calibration Pipeline scripts\footnote{ {\tiny\texttt{https://science.nrao.edu/facilities/vla/data-processing/pipeline}}}.  3C286 was observed as the primary flux density calibrator, the quasar PKS J1051$-$3138 was the complex gain calibrator, and 3C279 was the bandpass calibrator.  The absolute uncertainty in the overall flux density scale is estimated to be 10\,\%.  TW\,Hya exhibited significant proper motion over the timeframe of the observations (e.g, \citealt{2007A&A...474..653V}), and the extreme southern declination of this source relative to the latitude of the VLA may also contribute to the uncertainty in the absolute astrometry between observations, due to the uncertainty in the model of the atmospheric refraction in the correlator at high airmass.  The data obtained in the different VLA configurations were therefore successively aligned by matching the peaks in images from the next nearest VLA configuration, smoothed to the lower resolution to maximize the overlap in uv-coverage, before being combined. The reduced data were deprojected based on the inclination ($i=6^\circ$) and major-axis position angle ($\mathrm{PA}=335^\circ$) derived in \citet{2011ApJ...727...85H}, the same values that were used for deprojecting the SMA data \citep{2012ApJ...744..162A}. The deprojected imaginary components are effectively zero. The data in the different spectral bands are averaged to an effective wavelength of $9.315\,$mm, applying a spectral index of $\alpha=2.47$ (based on fitting the (sub-)mm tail of the spectral energy distribution, see also Sect.~\ref{sect:excess}). Fig.~\ref{fig:oldandnewVLA} shows a plot of the deprojected real component of the data.
 
Previous observations of TW\,Hya with the VLA in the sub-cm wavelength range were published by \citet{2007ApJ...664..536H} (see Sect.~\ref{sect:history}), in the epoch before VLA was upgraded. The observations are also shown in Fig.~\ref{fig:oldandnewVLA}, and suffer from a much lower signal-to-noise ratio than our new VLA data. In particular, we do not confirm the visibility null at $\sim1000\,$k$\lambda$, which was interpreted as being essential to confirm the disk structure with a 4-AU inner gap, proposed by \citet{2002ApJ...568.1008C}. 

TW\,Hya was also observed in the VLA A-configuration at 4.8\,GHz (6.3\,cm) and 7.3\,GHz (4.1\,cm) using the C-band receivers, with 1-GHz IFs (as for the Ka-band observations), in 2011 July. The VLA Calibration Pipeline scripts were used to calibrate and flag the data, and the uncertainty in the absolute flux density scale at these frequencies is estimated to be 5\,\%.  At a resolution of $1.0'' \times 0.3''$ (PA $\sim 15^\circ$) the source is unresolved, with peak flux densities of $75\pm7\,\mu$Jy and $145\pm21\,\mu$Jy, respectively.

\subsection{Spectral energy distribution}\label{sect:sed}
In addition to the multiwavelength interferometric data listed above, we compiled a spectral energy distribution (SED) for TW\,Hya. All data are listed in Table \ref{table:SED} with the corresponding references. As can be seen, the SED has a good coverage from the ultraviolet to mm-wavelengths. Far-UV data and cm-data are excluded, since emission at these wavelengths is affected by an accretion-related UV excess \citep{2010A&A...518L.125T} and a possible ionized wind \citep{2000ApJ...534L.101W,2012ApJ...751L..42P}, respectively. Emission at these wavelengths cannot be reproduced by the radiative-transfer code we will use. We will go deeper into this issue for cm-wavelengths in Sect.~\ref{sect:excess}.

For the actual model fits, we reset the minimum error on the photometry points to $5\,\%$ of the flux measured in the corresponding band.

The proximity of TW\,Hya ($d\sim50\,$pc) might imply that the photometric beam does not include the complete source flux. Due to the intrinsic spatial resolution and wavelength-dependency of the relevant emission, this effect can be expected to be strongest in the near-infrared, where a significant amount of scattered light might come from the outer disk. For both the IRAC and the WISE data, the beam sizes corresponding to the photometry effectively cover the whole target (see the references); for the 2MASS data, we note that the fraction of scattered light amounts to only $\sim2\,\%$, which is largely within the assumed $5\,\%$ accuracy for the SED fitting.

\begin{table}
\caption{Photometric fluxes for SED.}\label{table:SED}
\centering
{\small
 \begin{tabular}{lccl}
\hline\hline
photometric band & $\lambda_\mathrm{eff}$ ($\mu$m)&$F_\nu$ (Jy) & reference\\\hline
JOHNSON U & 0.36    & $0.050 \pm 0.005$ & 1 \\
TYCHO2 BT & 0.42    & $0.063 \pm 0.005$ & 2,3\\
JOHNSON B & 0.44    & $0.082 \pm 0.005$ & 1,4,5\\
HIPPARCOS HP & 0.52 & $0.140 \pm 0.005$ & 2\\
TYCHO2 VT & 0.53    & $0.116 \pm 0.008$ & 2,3\\
JOHNSON V & 0.55    & $0.155 \pm 0.009$ & 1,4,5\\
DENIS I & 0.79      & $0.428 \pm 0.008$ & 6\\
DENIS J & 1.23      & $0.87 \pm 0.04$ & 6\\
2MASS J & 1.24      & $0.81 \pm 0.02$ & 7\\
2MASS H & 1.65      & $0.99 \pm 0.04$ & 7\\
DENIS KS & 2.16     & $0.87 \pm 0.06$ & 6\\
2MASS KS & 2.16     & $0.81 \pm 0.02$ & 7\\
WISE W1 & 3.35      & $0.44 \pm 0.01$ & 8\\
IRAC 3.6 & 3.54     & $0.39 \pm 0.01$ & 9\\
CASPIR L & 3.58     & $0.39 \pm 0.02$ & 10\\
IRAC 4.5 & 4.48     & $0.267 \pm 0.004$ & 9\\
WISE W2 & 4.60      & $0.283 \pm 0.005$ & 8\\
IRAC 5.8 & 5.70     & $0.197 \pm 0.004$ & 9\\
IRAC 8.0 & 7.78     & $0.256 \pm 0.002$ & 9\\
AKARI S9W & 8.85    & $0.451 \pm 0.009$ & 11\\
IRAS F12 & 11.04    & $0.70 \pm 0.06$ & 12\\
WISE W3 & 11.55     & $0.445 \pm 0.006$ & 8\\
KECK 11.7 & 11.70   & $0.72 \pm 0.04$ & 13\\
KECK 17.9 & 17.90   & $1.45 \pm 0.08$ & 13\\
AKARI L18W & 18.92  & $1.45 \pm 0.04$ & 11\\
WISE W4 & 22.08     & $2.05 \pm 0.03$ & 8\\
IRAS F25 & 23.07    & $2.4 \pm 0.2$ & 12\\
MIPS 24 & 24.37     & $2.3 \pm 0.2$ & 14\\
IRAS F60 & 58.19    & $3.9 \pm 0.4$ & 12\\
MIPS 70 & 69.99     & $3.6 \pm 0.7$ & 14\\
PACS B & 70.39      & $3.9 \pm 0.2$ & 15\\
IRAS F100 & 99.52   & $5.0 \pm 0.5$ & 12\\
PACS R & 160.2     & $7.4\pm 0.7$ & 15\\
MIPS 160 & 161.6    & $6.6 \pm 1.3$ & 14\\
SHARC II & 350.0    & $6.1 \pm 0.7$ & 16\\
SCUBA 450WB & 456.4 & $4.3 \pm 0.9$ & 15\\
JCMT UKT14 & 800.0  & $1.45 \pm 0.03$ & 17\\
SCUBA 850WB & 858.9 & $1.4 \pm 0.1$ & 15\\
SMA CO3-2 & 867.0    & $1.46 \pm 0.04$ & 18\\
JCMT UKT14 & 1100   & $0.87 \pm 0.05$ & 17\\
SMA CO2-1 & 1300     & $0.57 \pm 0.02$ & 18\\
ATCA 3.4 & 3400     & $0.041 \pm 0.009$ & 19\\
VLA 7 & 7000        & $0.008 \pm 0.001$ & 20\\
\hline
 \end{tabular}}
\tablebib{
(1) \citet{2006yCat.2168....0M}; (2) \citet{1997ESASP1200.....P}; (3) \citet{2000A&A...355L..27H}; (4) \citet{2001KFNT...17..409K}; (5) \citet{2012yCat.5137....0A}; (6) \citet{2005yCat.2263....0D}; (7) \citet{2003yCat.2246....0C}; (8) \citet{2012yCat.2311....0C}; (9) \citet{2005ApJ...629..881H}; (10) \citet{2002ApJ...568.1008C}; (11) \citet{2010A&A...514A...1I}; (12) \citet{1988iras....7.....H}; (13) \citet{2002ApJ...566..409W}; (14) \citet{2005ApJ...631.1170L}; 
(15) \citet{2010A&A...518L.125T}; (16) \citet{2012ApJ...744..162A}; (17) \citet{1989ApJ...340L..69W}; (18) \citet{2004ApJ...616L..11Q}; (19) \citet{2003ApJ...596..597W}; (20) \citet{2000ApJ...534L.101W}.}

\end{table}

\section{Data inspection}\label{sect:inspection}
Before starting the effective modeling of the individual data sets, we start with a qualitative inspection of the different data sets.

\subsection{The lower limit of MIDI and a ring fit}\label{sect:ringfit}
TW\,Hya is at the faint end of objects that can be observed with MIDI. Fig.~\ref{fig:UV} gives an overview of all attempts to observe the target, showing both successful and unsuccessful fringe tracks. Inspecting the figure reveals that all observations on a projected baseline below $\sim50$\,m were successful, whereas observations above 50\,m failed. Unsuccessful attempts to start a fringe track can have two origins: \emph{extrinsic} factors like bad weather conditions (e.g.,~poor seeing) or instrumental failures (e.g.,~pointing problems, delay line problems); or \emph{intrinsic}, i.e.,~object-related, factors.

\begin{figure}
 \includegraphics[width=.5\textwidth,viewport=20 10 530 430,clip]{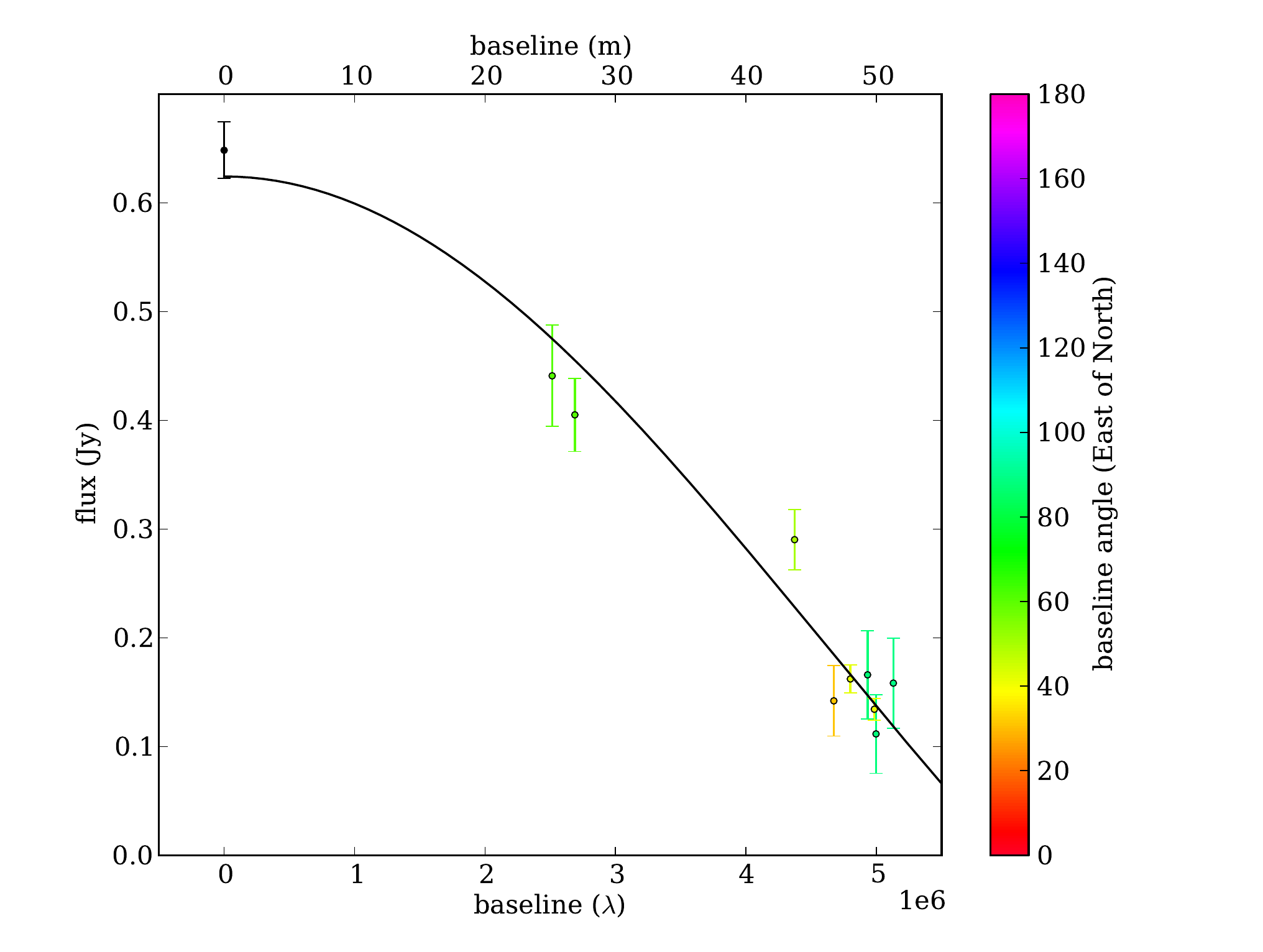}
 \caption{Correlated flux at $10\,\mu$m for each of the successful MIDI observations, color-coded following the baseline angle at which the observations were done. The baseline-0 observation (black point) corresponds to the average photometric flux of the observations. The full line shows a fit with the correlated-flux profile of an infinitesimally thin ring of angular diameter $\theta=26.3\,$mas (reduced $\chi^2=1.54$).}\label{fig:ring}
\end{figure}

Fig.~\ref{fig:ring} shows a plot of the correlated flux at 10\,$\mu$m vs.~the $uv$-distance, showing a uniform decrease in flux up to a baseline length of 52\,m, at a level of $\sim100$\,mJy. Although it can \emph{a priori} not be said whether the decrease in flux continues to longer baselines, it is likely that the unsuccessful observations are simply an effect of the correlated flux dropping below the sensitivity limit of MIDI (estimated to be $\sim80\,$mJy). Moreover, giving the nearly pole-on orientation of TW\,Hya, this effect should be symmetric, which is consistent with Fig.~\ref{fig:UV}. Hence, assuming that the unsuccessful observations are \emph{intrinsic} to the target, we find an upper limit of $\sim100$\,mJy for the $>50\,$m correlated fluxes.

To get a first constraint on the scale of the region emitting in the mid-infrared, we add a fit of a simple, infinitesimally thin ring model to Fig.~\ref{fig:ring}. The ring diameter is found to be $\theta=26.3\pm0.3\,$mas at 10\,$\mu$m (reduced $\chi^2=1.54$). At a distance of 51\,pc, this corresponds to a ring radius of $0.7\,$AU, similar to the scale found by \citet{2007A&A...471..173R}. Allowing the ring to be inclined, i.e.,~adding an inclination angle $i$ and a position angle PA as fit parameters, does not lead to a lower reduced $\chi^2$ ($\chi^2=1.75$). In other words, the MIDI observations are fully consistent with a face-on orientation of the disk.

\subsection{A scattering component}\label{sect:scattering}
Both the PIONIER and the NaCo/SAM data indicate that TW\,Hya is largely unresolved in the near-infrared, with baseline-averaged $V^2$ levels of $0.968\pm0.005$ and $0.949\pm0.006$, respectively (quoted errors are standard errors of mean). Roughly estimating the total flux as composed of a resolved and an unresolved component $F_\mathrm{resolved}$ and $F_\mathrm{unresolved}$, we have the following relation:
\begin{equation}
 F_\mathrm{resolved}=\frac{1-V}{V}\,F_\mathrm{unresolved}\label{eq:frac}
\end{equation}
($V=\sqrt{V^2}$ is the visibility amplitude). This shows that the fraction of resolved to unresolved near-infrared radiation is $2-3\,\%$ (at least, for the spatial ranges probed by the individual observations). 

The resolved component in the near-infrared can have two origins: thermal radiation originating from the hot inner-disk regions and/or stellar photons scattered in the disk atmosphere. Unlike thermal near-infrared radiation, which only originates from the hot inner region, scattered light will originate from the global disk. This translates into an immediate drop in visibilities on short baselines.

The azimuthally averaged SAM data clearly exhibit such a drop, revealing that at least a few percent of the L$'$-band flux is coming from large scales. The typical baselines on which this resolved flux is seen correspond to spacial scales of several tens of AU, confirming that the flux has a non-thermal---thus, scattering---origin. 

In the H-band, for the more noisy PIONIER data, the origin of the resolved flux is less clear. A useful point of additional information comes from Hubble-imaging by \citet{2002ApJ...566..409W}. These authors find a ratio of scattered to stellar flux of $0.021$ at $1.6\,\mu$m. The latter value inferred from large spatial scales ($R\gtrsim20\,$AU) is in good agreement with the $2\,\%$ resolved to unresolved flux ratio determined from the PIONIER data, for smaller scales ($\lesssim1\,$AU). This seems to indicate that the resolved flux in the H-band is predominantly coming from the outer disk regions, and hence has a scattering origin.   

The first near-infrared interferometric data of TW\,Hya were published by \citet{2006ApJ...637L.133E}. The wide-band visibility point at $2.14\,\mu$m valued $V^2=0.88\pm0.05$, for a 61.7-m baseline. In their modeling, the authors ignore the effect of scattering, and use the observation to constrain the inner-disk characteristics (see Sect.~\ref{sect:history}). The similar ratio of scattered to stellar flux found at $1.1-1.6$\,$\mu$m ($0.024-0.021$; \citealt{2002ApJ...566..409W}) and $3.6\,\mu$m ($0.027$; this paper) seems to indicate a roughly constant fraction of scattered flux across the range $1-4\,\mu$m. In other words, also in the K-band a contribution of $\sim2\,\%$ of scattered light to the total flux can be expected. A later reduction of the \citet{2006ApJ...637L.133E} data point put the visibility to $V^2=0.92\pm0.05$  (\citealt{2011ApJ...728...96A}; these authors account for a flux bias). Taking this and Eq.~(\ref{eq:frac}) into consideration, the K-band visibility point might not be related to thermal emission, but purely reflect the scattered flux in the K band. \citet{2011ApJ...728...96A} reach a similar conclusion, where new K-band interferometry confirms an overresolved component.

\subsection{Symmetry and variability}
As observations tend to be more detailed, protoplanetary disks are revealed to possess more asymmetrical features than originally thought, in some cases very spectacular (e.g.,~\citealt{2011ApJ...729L..17H,2013ApJ...762...48G,2013Natur.493..191C}). Also variability is found to be not restricted to the optical and near-infrared spectral range only (e.g.,~\citealt{2012ApJ...744..118J,2012arXiv1203.6265P}). For TW\,Hya, indications for disk asymmetries and variable disk emission are relatively limited. \citet{2005ApJ...622.1171R} report a sinusoidal dependence in the optical surface brightness on the azimuth angle, and suggest this as an indication for a warped inner disk. In the mid-infrared, \citet{2012ApJ...750..119A} model speckle-interferometry data with a disk and a hotter-than-equilibrium companion. These authors also indicate variability in the 11.6-$\mu$m emission of two epochs. Multi-epoch Spitzer observations might also indicate a variable flux level in the mid-infrared, although pointing errors cannot be excluded for these observations \citep{2010ApJ...712..274N}.

In both our infrared and (sub-)mm data, no significant asymmetries are found. The closure phases in the SAM data were investigated in detail in \citet{2012ApJ...744..120E}, but no departures from point symmetry were detected. The deprojected (sub-)mm data show an imaginary component that is effectively zero, also indicating full circular symmetry. Finally, considering the variability, our multi-epoch 50-m baseline observations on MIDI have fully consistent flux levels. Our observations thus do not confirm the variability seen in the mid-infrared based on single-dish observations (at least, within the $\sim10\,\%$ calibration accuracy of MIDI).

\section{Reimplementing key models}\label{sect:reimplementing}
Although we could immediately start modeling the new data sets with a custom radiative transfer code, it is instructive to take an intermediate step. As was mentioned above, several disk models have been published throughout the last decade. Since these models are based on different modeling techniques, it is important to distinguish between the essential characteristics of the models and the model-intrinsic parameters that are less relevant. To achieve this, we make a reimplementation of a selection of models discussed in Sect.~\ref{sect:history}.

\subsection{Three key models}
We select three models that were of a key importance for our understanding of the structure of the TW\,Hya disk. 
These models are from:
\begin{itemize}
\item \citet{2002ApJ...568.1008C}: this is the ``gap-discovery'' paper, with a detailed argumentation for the dust populations;
\item \citet{2007A&A...471..173R}: this is the first work which argues for a different inner-disk structure, with a smaller inner gap ($R_\mathrm{in}\sim0.7\,$AU rather than 4\,AU);
\item \citet{2012ApJ...744..162A}: this paper presents a direct measurement of the disk size at mm-wavelengths, rather different than the scattered-light edge ($R_\mathrm{out}\sim60\,$AU rather than $>200\,$AU).
\end{itemize}
These models distinguish themselves from the others since they are radiative-transfer models for the complete disk (with realistic dust mixtures), and since they provided new structural constraints that have been essential for works that followed on these respective publications. As an example, the works by \citet{2006ApJ...637L.133E} and \citet{2007ApJ...664..536H} have taken the Calvet model as a starting model, while \citet{2011ApJ...728...96A} have essentially combined properties of the Calvet and the Ratzka models.

We will refer to the three models as the ``key'' models, and refer to the individual models by the first-author's name.

\begin{figure}
 \includegraphics[width=.5\textwidth,viewport=10 5 400 340,clip]{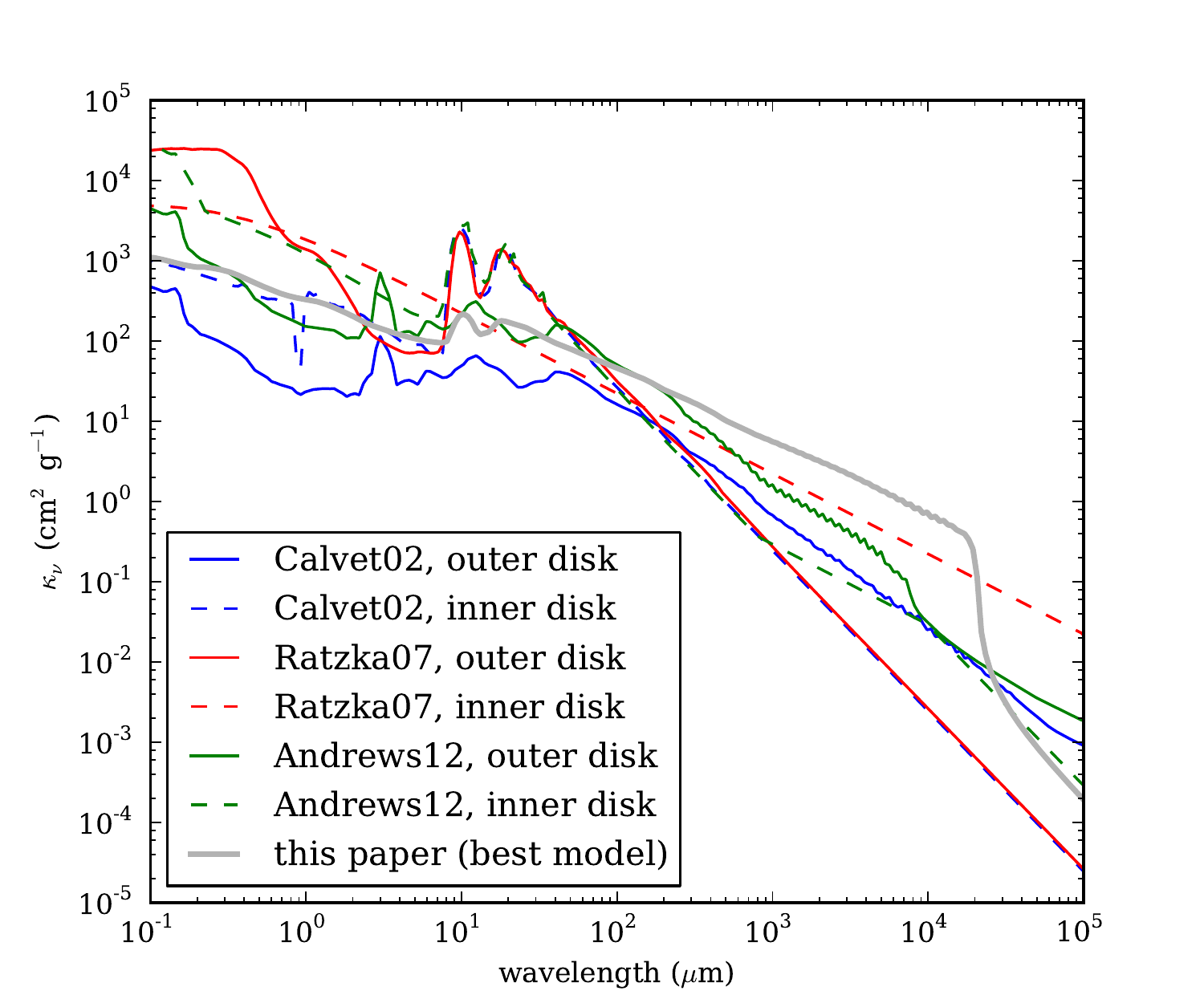}
 \caption{Comparison of the mass absorption coefficients of the dust mixtures in the different reimplemented models, distinguishing between the inner (\emph{dashed}) and outer (\emph{full}) disks. In gray, we show the dust mixture of the final best model (Sect.~\ref{sect:modeling}).}\label{fig:modelopacities}
\end{figure}

\subsection{The radiative transfer code}
Since there have been no indications for strong azimuthal asymmetries in the disk around TW\,Hya\footnote{The hint for asymmetry in \citet{2012ApJ...750..119A} is interpreted as an indication for a (unconfirmed) secondary source, and \emph{not} as an indication for disk asymmetries.}, it is appropriate to start with an azimuthally symmetric disk model for simulating the data. We will use the Monte Carlo radiative transfer code \texttt{MCMax} \citep{2009A&A...497..155M} to address the full radiative transfer problem. This code has recently been used to make a coherent picture of several protoplanetary-disk objects (e.g.,~for HD\,100546, \citet{2011A&A...531A..93M}; HD\,142527, \citet{2011A&A...528A..91V}; HD\,169142, \citet{2012ApJ...752..143H}).

In \texttt{MCMax}, an azimuthally symmetric disk structure is set up and, if specified, the code solves for vertical hydrostatic equilibrium by iterating on the temperature and density structures (e.g.,~\citealt{2007prpl.conf..555D}).  A recent update to the code followed on the work of \citet{2012A&A...539A...9M}, and implements vertical settling based on the $\alpha$-disk turbulence prescription. For high turbulence strengths $\alpha$, grains will experience a strong gas turbulence, and even large grains couple to the gas. Lowering $\alpha$ leads to the decoupling of the larger grains (because of there low surface-to-mass ratios): these grains settle to the midplane.

For all the models, we assume dust particles to be thermally coupled and gas temperatures to be set by the dust temperatures.

\subsection{Model reimplementation}
We list the disk parameters and a graphical representation of the three key models in Table \ref{table:reimplementation}. Each model conceptually has an inner and an outer disk, which implies we could split the table in two. For each model, we carefully reassembled the dust mixtures used in the papers, and traced back the optical constants from the original papers (Fig.~\ref{fig:modelopacities}).

Using a versatile radiative-transfer code, it is possible in principle to fully reconstruct the models with their different particularities. Each of the three models has a different basic model to start from: the Calvet model is based on the \citet{1998ApJ...500..411D} disk model, the Ratzka model uses the disk formalism of \citet{1997ApJ...490..368C}, and the Andrews model makes use of a genuine Monte Carlo radiative transfer code (\texttt{RadMC}, \citealt{2004A&A...417..159D}). The different model structures complicate the model reconstruction in two ways:
\begin{itemize}
 \item vertical disk structure: each model has its own formalism for calculating the vertical disk structure (e.g.,~hydrostatic equilibrium vs.~parameterized);
 \item disk rim: in each of the three models, the shape of the (inner) rim of the outer disk seems to be an important fine-tuning parameter for getting the right flux in the mid-infrared. 
\end{itemize}
We decide to keep the radial structure of the models, but homogenize the vertical structure. For this, we make use of the ability of the radiative-transfer code \texttt{MCMax} to calculate self-consistently vertical hydrostatic equilibrium, assuming the dust and gas are well coupled. However, we do allow for a scaling factor $\Psi$ between the vertical dust scale height and the gas scale height, and reconstruct the shape of the disk rim.

For reference, Table \ref{table:reimplementation} lists to which data the original models were fit. This allows us to interpret the quality of the model reimplementation below in terms of the original model capacities.

\begin{table*}[!t]
 \centering
\caption{Parameters taken for model reimplementation. Essentially all parameter values are exact copies of the original papers. }\label{table:reimplementation}
\begin{tabular}{lrrr}
 \hline\hline &Calvet et al.~(2002) & Ratzka et al.~(2007) & Andrews et al.~(2012)\\
 \hline \emph{Figure}\\&\includegraphics[width=0.2\textwidth]{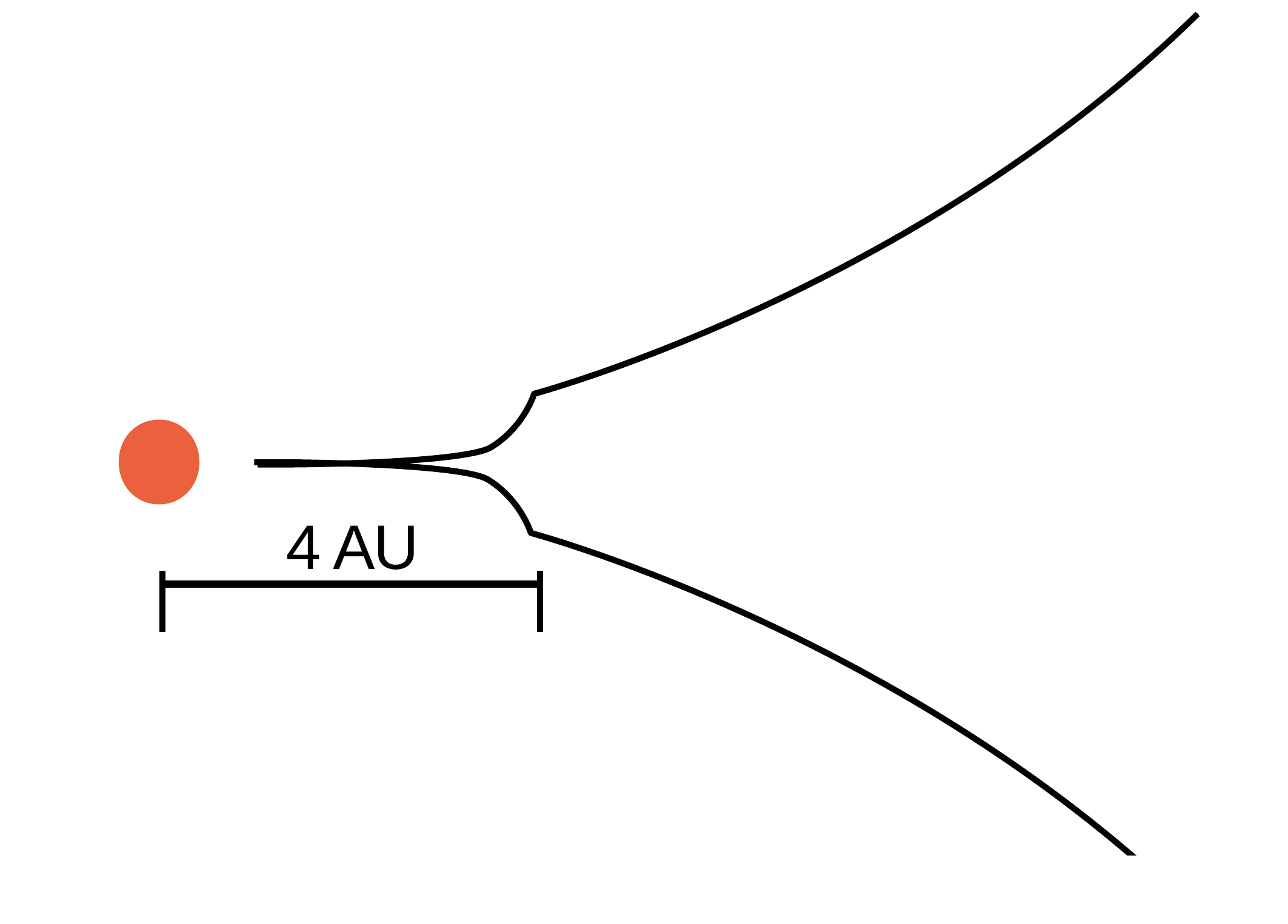}& \includegraphics[width=0.2\textwidth]{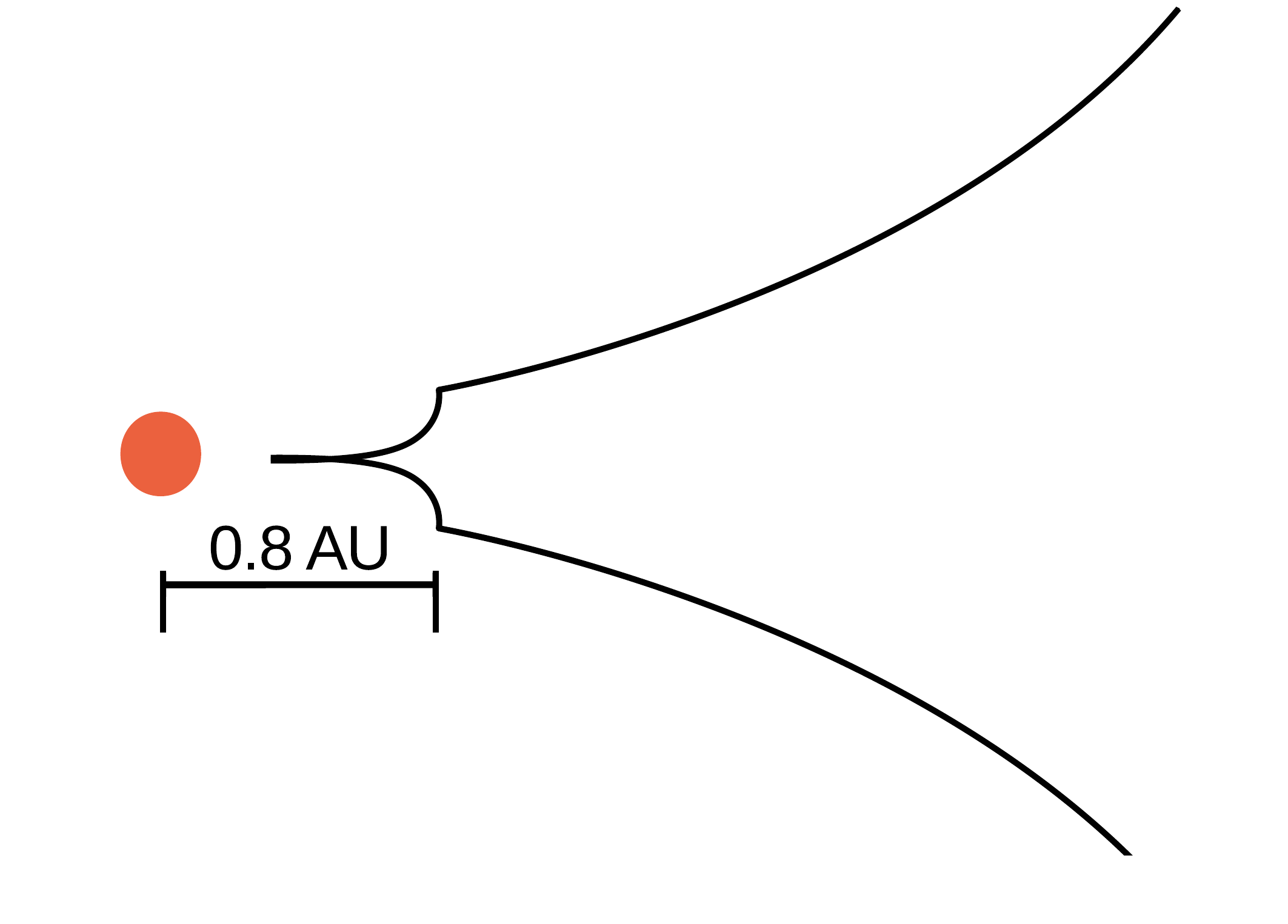}& \includegraphics[width=0.2\textwidth]{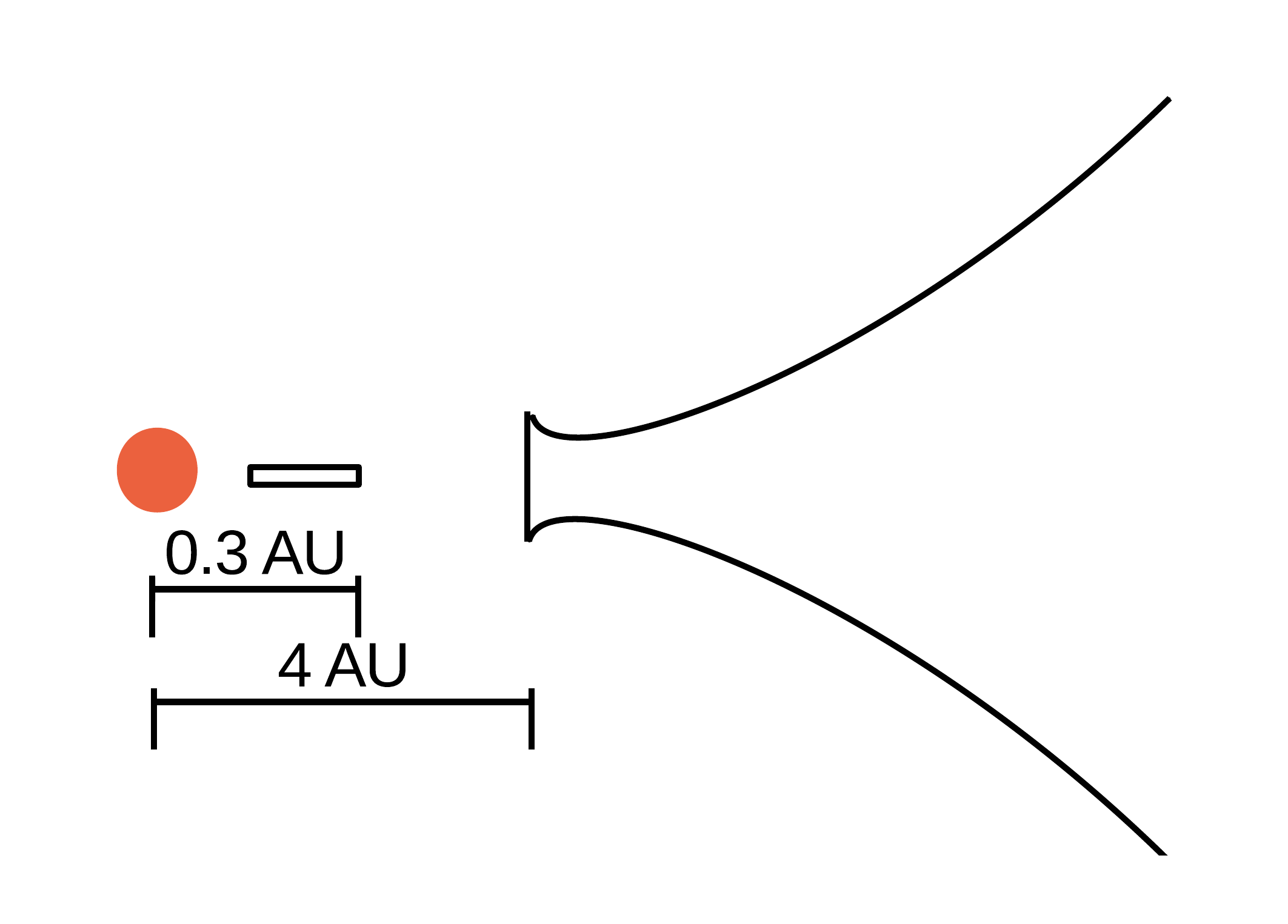}
\\\hline\emph{Originally fit data}&SED&SED+MIDI&SED+SMA\\\hline\emph{Inner disk} &&&\\
 inner radius $R_\mathrm{in}$ (AU)&0.02&0.06&0.05\\
 outer radius $R_\mathrm{out}$ (AU)&3&0.5&0.3\\
 dust mass $M_\mathrm{dust}$ ($\mathrm M_\odot$)&$2.0\times10^{-8}$&$6\times10^{-13}$&$1.55\times10^{-11}$\\
 surf.~dens.~power $p$&0&1.5&0\\
composition\tablefootmark{a}&glassy pyroxene\tablefootmark{b} &$\lambda^{-1}$-law\tablefootmark{c}  &mixture\tablefootmark{d}\\
&(100\,\% $0.9-2\,\mu$m)&(100\,\%)&(75\,\% $0.005-1\,\mu$m)\\
&&&forsterite\tablefootmark{e}\\
&&&(25\,\% $0.005-1\,\mu$m)\\
\hline \emph{Outer disk}&&&\\
 inner radius $R_\mathrm{in}$ (AU)&$3-4$&$0.5-0.8$&4\\
 outer radius $R_\mathrm{out}$ (AU)&140&140&60\\
 dust mass $M_\mathrm{dust}$ ($\mathrm M_\odot$)&$7.8\times10^{-4}$&$1.6\times10^{-3}$&$2.0\times10^{-4}$\\
 surf.~dens.~power $p$&1&1.5&0.75\\
settling $\Psi$&1&0.7&1.3\\
composition\tablefootmark{a}&mixture\tablefootmark{f} &silicate~mixture\tablefootmark{g} &mixture\tablefootmark{d}\\
&(100\,\% $0.005-10^{4}\,\mu$m)&(96\,\% $0.1,1.5,6\,\mu$m)&(95\,\% $0.005-10^3\,\mu$m)\\
&&carbonaceous grains\tablefootmark{h}&mixture\tablefootmark{d}\\
&&(4\,\% $6\,\mu$m)&(5\,\% $0.005-1\,\mu$m)\\
rim composition&same&same&mixture\tablefootmark{d}\\
&&&(100\,\% $0.005-1\,\mu$m)\\
rim shape&flaring&flaring&puffed-up vertical wall\\
\hline\\
\end{tabular}
\tablefoot{
\tablefoottext{a}{In mass fractions. All size distributions use $-3.5$ as the particle-size density power law exponent, and extinction coefficients are calculated for spherical grains and/or distributions of hollow spheres (DHS, \citealt{2005A&A...432..909M}), following the choices in the original papers.}\\
\tablefoottext{b}{\citet{1994A&A...292..641J}}\\
\tablefoottext{c}{$\,\kappa_\lambda(2\,\mu\mathrm{m})=10^3$\,cm$^2$\,g$^{-1}$}\\
\tablefoottext{d}{amorphous silicates (24\,\%; \citealt{1984ApJ...285...89D}), FeS (6\,\%; \citealt{1994ApJ...423L..71B}; see \citealt{1999A&AS..136..405H}), organics (30\,\%; \citealt{1994ApJ...421..615P}), water ice (40\,\%; \citealt{1984ApOpt..23.1206W})}\\
\tablefoottext{e}{\citet{1973PSSBR..55..677S}}\\
\tablefoottext{f}{amorphous silicates (24\,\%; \citealt{1994A&A...292..641J}), FeS (6\,\%; \citealt{1994ApJ...423L..71B}; see \citealt{1999A&AS..136..405H}), organics (30\,\%; \citealt{1994ApJ...421..615P}), water ice (40\,\%; \citealt{1984ApOpt..23.1206W})}\\
\tablefoottext{g}{amorphous silicates: olivines MgFeSiO$_4$ (66\,\% 0.1\,$\mu$m, 2\,\% 6.0\,$\mu$m; \citealt{1995AA...300..503D}),
pyroxenes MgFe(SiO$_3$)$_2$ (18\,\% 0.1\,$\mu$m, 3\,\% 1.5\,$\mu$m; \citealt{1995AA...300..503D}); crystalline silicates: forsterite (3\,\% 0.1\,$\mu$m; \citealt{1973PSSBR..55..677S}), enstatite (4\,\% 0.1\,$\mu$m, 1\,\% 1.5\,$\mu$m; \citealt{1998A&A...339..904J})}\\
\tablefoottext{h}{\citet{1993A&A...279..577P}}}
\end{table*}

\begin{figure*}
 \includegraphics[width=.49\textwidth,viewport=0 10 532 410,clip]{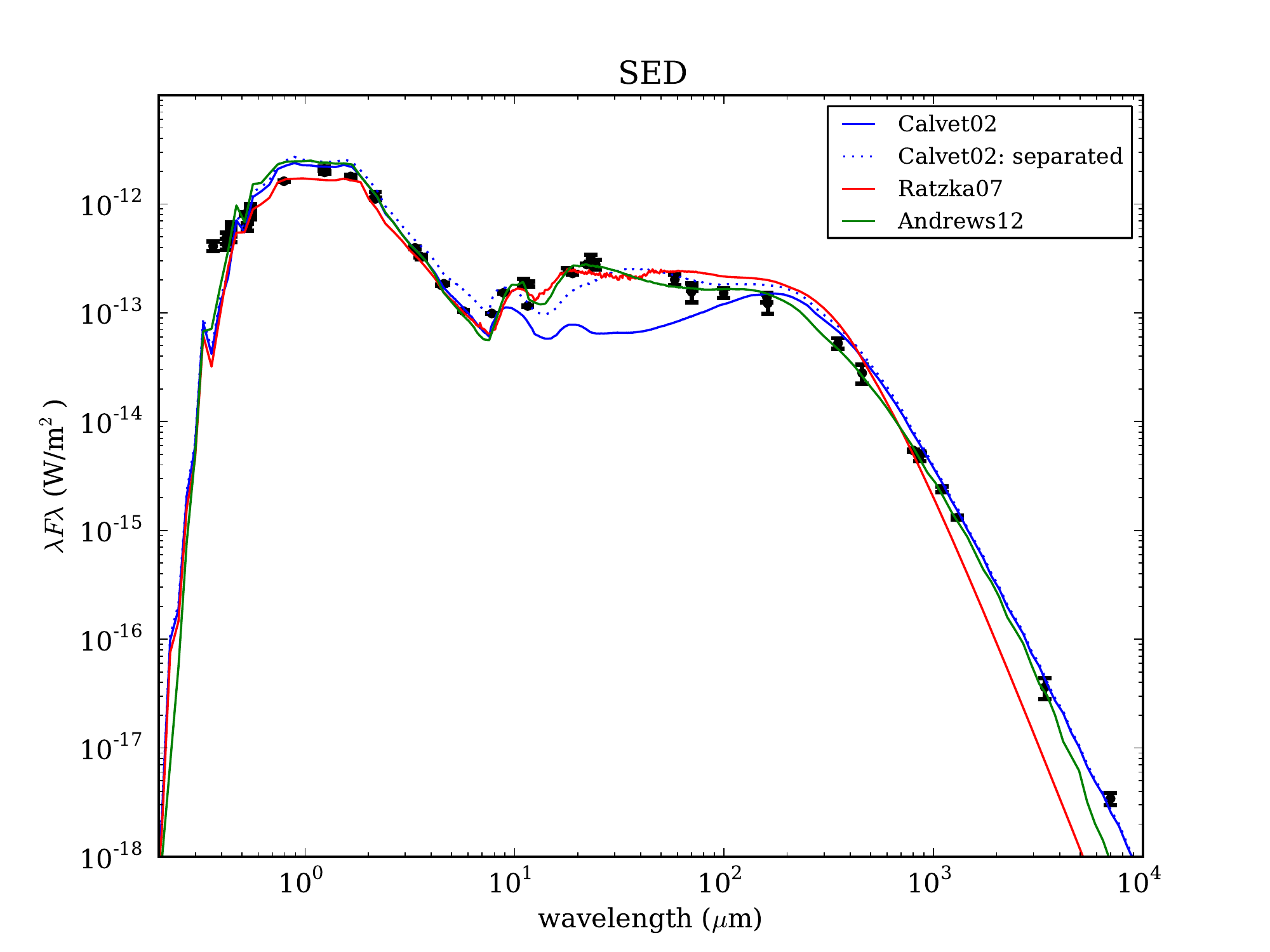}
 \includegraphics[width=.49\textwidth,viewport=0 10 532 410,clip]{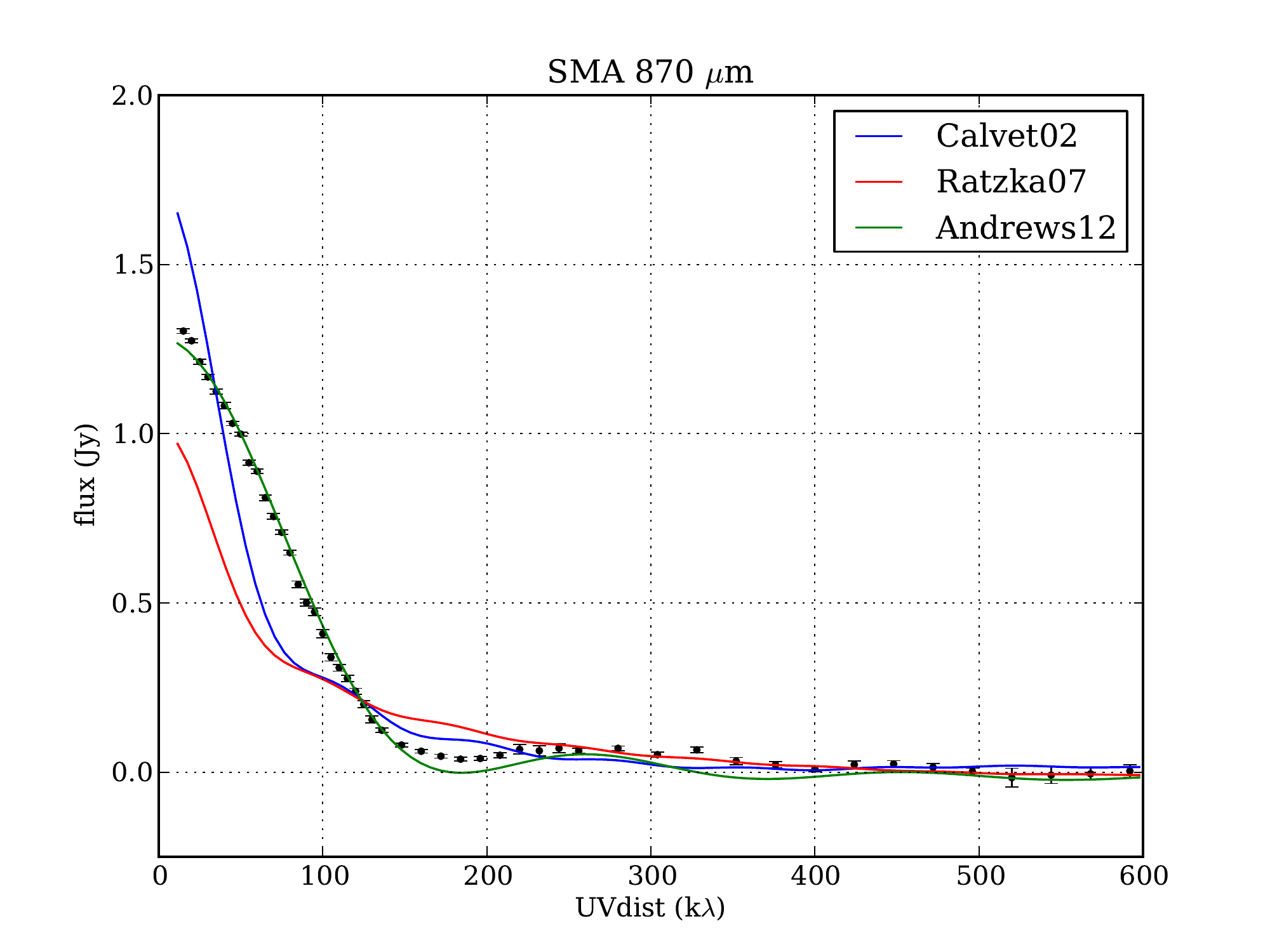}\\
 \includegraphics[width=.49\textwidth,viewport=0 10 532 410,clip]{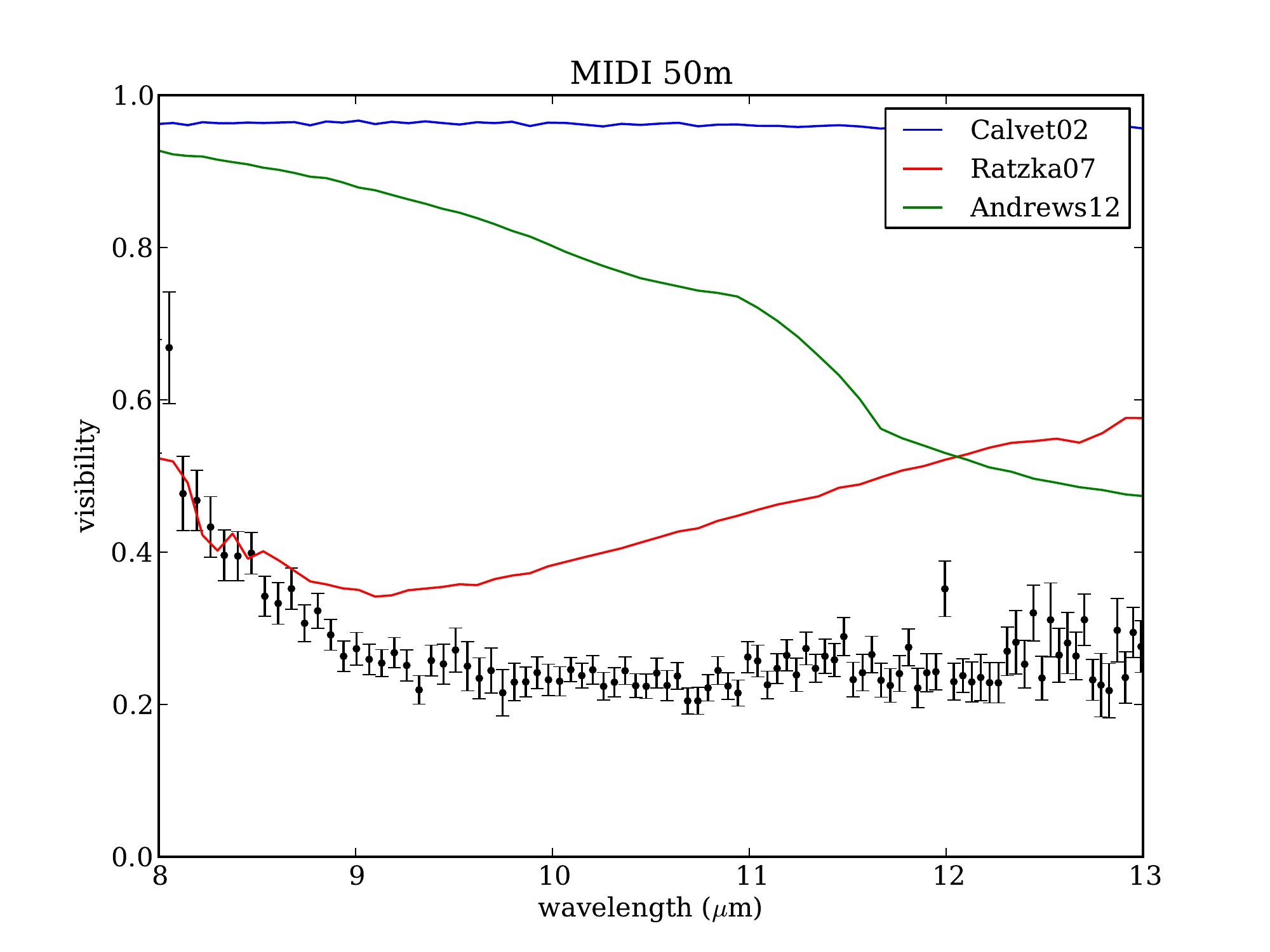}
 \includegraphics[width=.49\textwidth,viewport=0 10 532 410,clip]{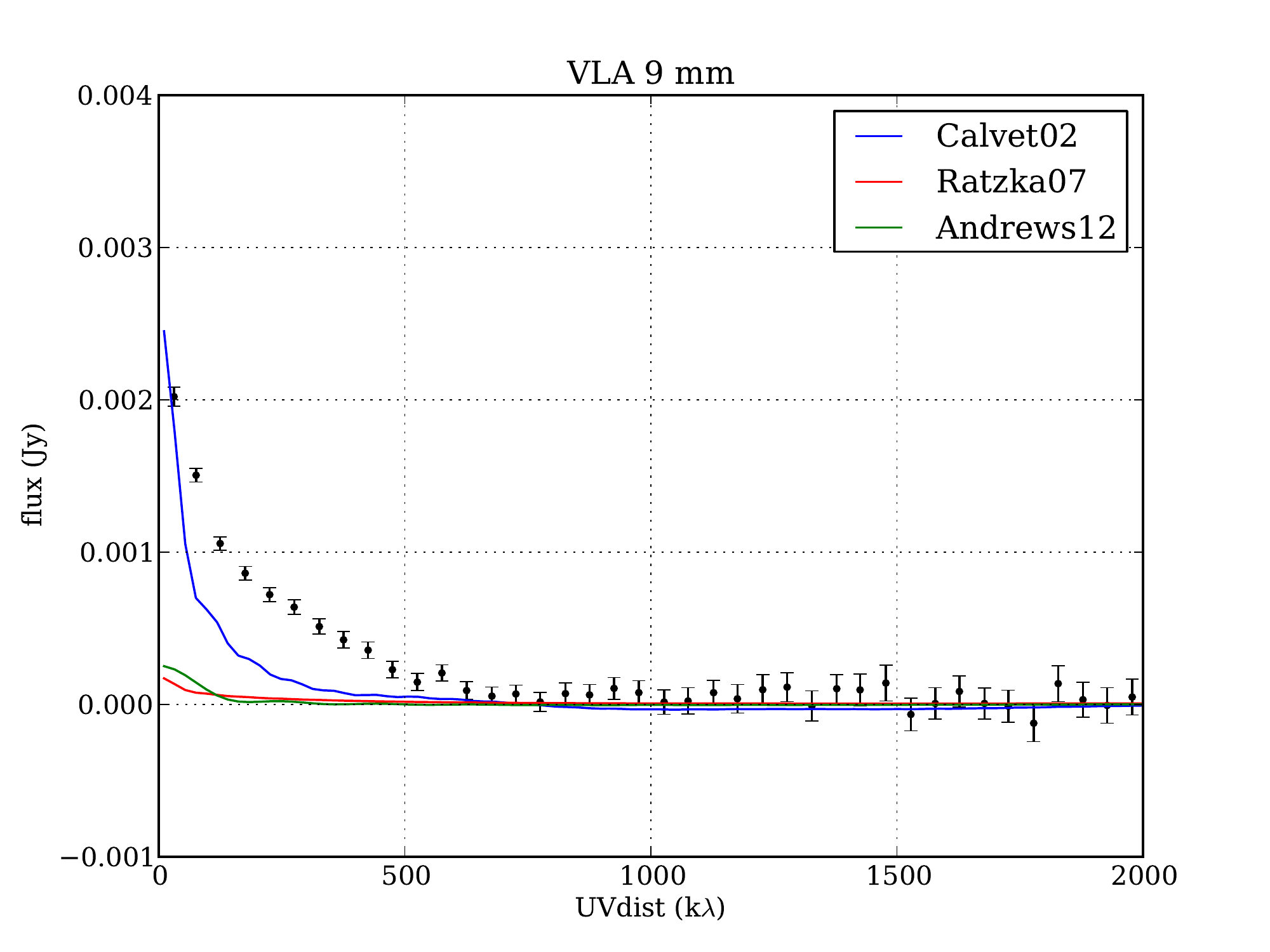}
\caption{Comparison of the SED data, the MIDI 50-m visibility profile, the deprojected SMA visibilities, and the deprojected VLA visibilities with the reimplemented model data. Along with the fully reimplemented \citet{2002ApJ...568.1008C} model SED, we show the SED consisting of the sum of the contributions of the inner-disk and the outer-disk SED, for which the radiative transfer is done separately.}\label{fig:reimplementation}
\end{figure*}

\subsection{Results}
In Fig.~\ref{fig:reimplementation}, the resulting SED of each model reimplementation is shown. Since all three original models reproduce the SED well, any difference between the reimplemented model and the SED is at least partly due to a departure in the intrinsic model structure.  

Our reimplementation of the Andrews model accurately reproduces the SED from optical to sub-mm wavelengths. We therefore believe it to be a valid reconstruction of the original model. Like the model reproduction, the original model is constructed with a Monte Carlo radiative transfer code. To imitate the vertical structure, which is parametrized in the original model, we scaled the self-consistent hydrostatic-equilibrium scale height by $\Psi=1.3$. At mm-wavelengths, the model flux falls below the observed fluxes. This regime is not constrained by the original model, and possibly reflects the need for an alternative large-grain population (e.g.,~cm-sized grains). 

The reimplementation of the Calvet and the Ratzka models is more problematic. As can be seen in Fig.~\ref{fig:reimplementation}, the former SED has a strong lack of mid- and far-infrared flux, whereas the latter is fitting poorly at (sub-)mm wavelengths. Both models have been produced based on disk-model formalisms preceding state-of-the-art Monte Carlo radiative-transfer modeling. The original models thus lack the self-consistent temperature calculation. In the case of the Calvet model, although the inner disk is relatively tenuous, it is strongly optically thick in the radial direction. The shadowing effects, which could not be taken into account in the original model, cause a strong lack in flux from the mid- to the far-infrared. To demonstrate the effect, we also show the sum of the SEDs made by modeling the radiative transfer of the inner and outer disk separately. The resulting artificial SED clearly reproduces the photometry more accurately, confirming the nature of the original Calvet model. For the Ratzka model, the model reimplementation fits relatively well in the mid-infrared. Here, the inner disk is much more tenuous, and shadowing effects are not playing an important role. The departures of the spectral index at (sub-)mm wavelengths reflects the lack of grains larger than micron-sizes. The latter grains are essential for reproducing the mm-slopes generally seen in protoplanetary-disk SEDs (e.g.,~\citealt{2007prpl.conf..767N}). The original model seems not to indicate difficulties in fitting the slope (see Fig.~7 in \citet{2007A&A...471..173R}), but our full radiative transfer clearly shows this regime to be problematic. We therefore interpret this departure as a clear indication for a large-grain population.

In Fig.~\ref{fig:reimplementation}, the mid-infrared visibilities on a 50-m baseline are also shown, for each model reproduction (we limit the comparison to the 50-m visibilities for clarity of the plot). Comparing these profiles to the corresponding MIDI data shows that the Ratzka model has the best overall agreement, although a significant departure is seen at wavelengths longer than 10\,$\mu$m. In both the Calvet and Andrews model, the mid-infrared radiation is coming from a compact inner region: the central part of the optically-thin inner dust region ($R_\mathrm{in}=0.02\,$AU) and the dust ring ($R=0.05-0.25$\,AU), respectively. In the Ratzka model, the 10-$\mu$m feature is associated with the disk rim ($R_\mathrm{in}=0.5\,$AU), and very little mid-infrared emission is coming from within this rim. The original model was fitting the (originally reduced) data relatively well, but the new data reduction has lowered the visibilities at wavelengths above 10\,$\mu$m. This largely explains the departure.

In the sub-mm regime measured by SMA, the Andrews model is superior, well reproducing the visibility lobes from short to long baselines (Fig.~\ref{fig:reimplementation}). The outer disk in the other two models is much larger ($R_\mathrm{out}=140\,$AU vs.~60\,AU for the Andrews model), explaining the steeper drop in visibility amplitude at short baselines. Clearly, the Andrews model was designed to fit the SMA data, and we accurately reimplemented the original model.

Finally, none of the three model reimplementations can reproduce the VLA 9-mm data. For the Ratzka and the Andrews model, the lack of mm-emission (seen in the SED) translates into a visibility level much lower than the one observed. In the Calvet case, we should be more careful in making conclusions. In principle, the original Calvet model should agree to some extent with the mm-visibilities, since this model (in a slightly adapted version) was shown to agree with the original VLA data of TW\,Hya (see \citealt{2007ApJ...664..536H}). However, we already indicated in Sect.~\ref{sect:mminterfero} that the signal-to-noise of those data was much lower than that of the new VLA data presented here. The new data thus put new constraints on the models, which were not available before, and could therefore explain possible departures. However, we believe most of the disagreement between our model reimplementation and the mm-data to be related to the intrinsic model characteristics. Using our full Monte Carlo radiative transfer implementation of the Calvet model, we were unable to obtain similar mm-visibility profiles as the original model. Irrespective of the quality of the previous and the new data, it seems thus impossible to reproduce the bright inner rim in the original semi-analytical model in \citet{2007ApJ...664..536H}.

\subsection{Essential features}
The results of the model reimplemenation can now be interpreted in terms of a few essential features which need to be investigated in detail. 

\paragraph{Different model philosophies.} That different models with strong structural differences reproduce the SED (to some extent) reconfirms that SED modeling is at least partly degenerate. In particular, it does not allow the distinction between two different model ``philosophies'': a large-gap disk with a 10-$\mu$m excess coming from an optically thin inner region (Calvet and Andrews model), or a small-gap disk with its own silicate feature, with some extra continuum-opacity source inside (Ratzka model). 

\paragraph{Grain properties.}
\citet{2012ApJ...750..119A} argue that the differences in mid-infrared opacities of the grain species taken for the Calvet model, on the one hand, and the Ratzka model, on the other, explain why a different hole size ($4\,$AU vs.~$<1\,$AU) is found. We question this reasoning because:
\begin{itemize}
 \item The differences in the mid-infrared opacities are marginal compared to the ones at other wavelengths, a wavelength-range which is not considered by \citet{2012ApJ...750..119A}. Also, the mid-infrared opacities of the Andrews model differ very similarly (like the Ratzka model), yet these authors also have a 4-AU gap.
 \item The 10-$\mu$m feature in the Ratzka and in the 4-AU gap models do not have the same origin: in the former, radiation is coming from the outer-disk rim, in the latter, from the inner optically thin disk. Therefore, it is incorrect to consider the Ratzka model as a ``scaled'' 4-AU gap model, i.e.,~as the same model with a smaller inner hole.
\end{itemize}
Considering the maximum grain size, the SEDs have shown that large grains are needed to reproduce the mm-slope. 

\paragraph{Full radiative transfer.}
Our reimplementation of the three models has shown that fully solving the radiative transfer in a disk is important (the same conclusion concerning the inclusion of scattered light is drawn by \citealt{2008ApJ...673L..63P}). Structures that are optically thin in the vertical direction are not necessarily optically thin along the radial direction, and can therefore cast a shadow on the outer disk. The loss in intercepted stellar radiation decreases the thermal emission coming from the corresponding scales. This is essentially what was demonstrated for the Calvet model, where the mid-infrared flux to a large extent vanished when self-consistently calculating the model structure.

Solving the full radiative-transfer problem also allows us to use the spectral signature of the small-grain silicates as an important diagnostic for the fit quality. Using simple powerlaw opacity laws (e.g.,~\citealt{2006ApJ...637L.133E}, \citealt{2011ApJ...728...96A}) to some extent misses the interpretation in terms of realistic particle characteristics (e.g.,~composition, grain sizes).

\paragraph{To 4 AU or not to 4 AU?}
Perhaps the most interesting parameter for the TW\,Hya disk is the size (or size range) of the inner gap. As we have shown, the current 4-AU gap models do not manage to model the MIDI visibilities, and also the new VLA data seem not to confirm the essential visibility null found in the previous VLA data. However, the data presented in \citet{2012ApJ...750..119A} still are in agreement with a 4-AU inner gap, so some structure change on this scale still might be needed for explaining all data.

\section{Radiative transfer modeling}\label{sect:modeling}
The inability of the previous models to fully explain the combined data set confirms the need for the development of a new disk model. This is what is done in the current section. 

\subsection{Model fitting strategy: a genetic algorithm}
We decided to use a genetic fitting algorithm for the radiative transfer modeling of the spectrally and spatially resolved data. This fitting strategy allows for an efficient exploration of the parameter space and thereby limits the risk of getting stuck in local $\chi^2$-minima. 

The genetic-algorithm formalism used for the model fitting in this work is described in Appendix \ref{appendix:GA}.

\begin{figure*}
 \centering
\includegraphics[width=.8\textwidth]{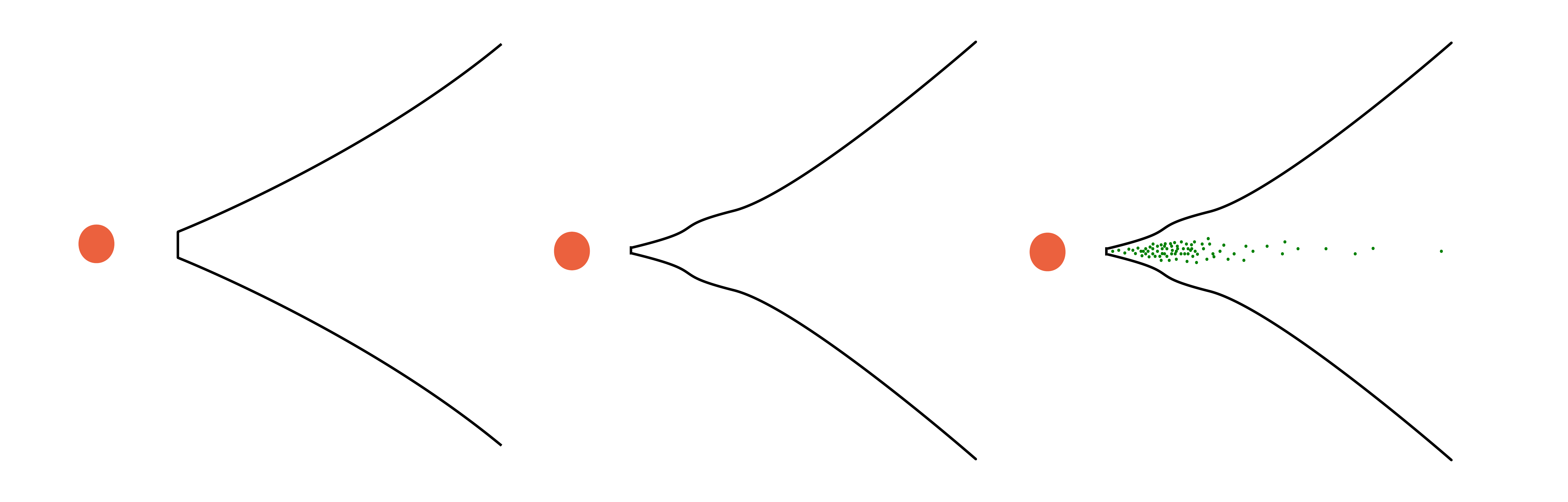}
\caption{Model refinement strategy: we start with a \emph{simple disk} with a vertical inner rim (\emph{left}), we then introduce a \emph{rounded inner rim} (\emph{middle}), and finally also \emph{concentrate the largest grains towards the inner disk regions} (\emph{right}).}\label{fig:refinement}
\end{figure*}

\subsection{Stellar properties and dust composition}
In principle, the complete set of parameters involved in a disk model could be included into the fitting process. However, even using a dedicated fitting strategy, we have a strong interest to minimize the actual fitting parameters, speeding up and mitigating the fitting process. 

During the radiative-transfer modeling, the stellar properties are kept fixed to literature values: stellar mass $M_\star=0.6\,$M$_\odot$, luminosity $L=0.25\,$L$_\odot$, effective temperature $T_\mathrm{eff}=4000\,$K, distance $d=51\,$pc (references in \citet{2007A&A...471..173R}; the luminosity was slightly increased). We also fix the disk inclination to $i=0^\circ$: we model the deprojected visibility curves in the (sub-)mm, and the relatively poor uv-coverage and/or the error level of our infrared interferometry makes the 6-degree inclination correction irrelevant.

For the dust composition we use a mixture of amorphous silicates (Table \ref{table:amsil}) and carbonaceous grains. Opacities are calculated using a distribution of hollow spheres (DHS; \citealt{2005A&A...432..909M}) for the particle shapes. To model the silicate composition, we use the diagnostics of the 10-$\mu$m feature in a Spitzer IRS spectrum of TW\,Hya (kindly provided by T.~Ratzka; see \citealt{2007A&A...471..173R}), using a procedure similar to the one in, e.g.,~\citet{2005A&A...437..189V} and \citet{2010ApJ...721..431J}. The contribution of the dust to the spectrum is fit using the following equation:
\begin{equation}
F_\nu=B_\nu\big(T_\mathrm{warm}\big)\,A_\mathrm{warm}
+B_\nu\big(T_\mathrm{cold}\big)\,A_\mathrm{cold}+ B_\nu\big(T_\mathrm{dust}\big)\sum_i\kappa_{\nu,i}\,A_i,
\end{equation}
where $\kappa_{\nu,i}$ is the dust opacity of silicate species $i$, and $A_x$ are the weighting factors of the individual terms. In words, we assume the dust emission to be composed of the thermal (continuum) emission of two black-body components ($B_\nu$) at characteristic temperatures $T_\mathrm{warm}$ and $T_\mathrm{cold}$, and optically thin emission of the dust grains at a characteristic dust temperature $T_\mathrm{dust}$.

\begin{table}
\centering
\caption{Amorphous silicates used for the spectral fit.}\label{table:amsil}
 \begin{tabular}{lll}
 \hline\hline
  dust species\tablefootmark{a}&chem.~formula&reference\\\hline
  olivine&Mg$_2$SiO$_4$&\protect\citet{2003AA...408..193J}\\
   &MgFeSiO$_4$&\protect\citet{1995AA...300..503D}\\
   pyroxene&MgSiO$_3$&\protect\citet{2003AA...408..193J}\\
   &MgFe(SiO$_3$)$_2$&\protect\citet{1995AA...300..503D}\\\hline
 \end{tabular}
\tablefoot{
\tablefoottext{a}{All opacities calculated using DHS; particle sizes $a=0.01-10^5\,\mu$m, size distribution $n(a)\propto a^{-3.5}$.}}
\end{table}

The best fit to the Spitzer spectrum is reached for a mixture of pure amorphous magnesium-iron olivines (MgFeSiO$_4$). Our final dust mixture is taken to be $80\,\%$ of these silicates and $20\,\%$ of carbonaceous grains \citep{1993A&A...279..577P}, the latter representing the continuum opacity. A constant dust-to-gas ratio of $0.01$ is assumed. Following our Spitzer-spectrum fit, we also fix the index of the particle size distribution to $-3.5$ \citep{1977ApJ...217..425M} and the minimum particle size to $a_\mathrm{min}=0.01\,\mu$m (however, the maximum grain size $a_\mathrm{max}$ will be varied; see Sect.~\ref{sect:simpledisk}).

\begin{table}
\centering
\caption{Parameter ranges for disk models.}\label{table:single}
 \begin{tabular}{lll}
  \hline\hline parameter & range/values & sampling\tablefootmark{a} \\\hline
 inner radius $R_\mathrm{in}$ (AU)&$[0.05,1]$&lin.\\
 outer radius $R_\mathrm{out}$ (AU)&$[30,300]$&lin.\\
 dust mass $M_\mathrm{dust}$ (M$_\odot$)&$[10^{-4.5},10^{-2.5}]$&log.\\
 surfdens.~power $p$&$[0,2]$&lin.\\
 max.~grain size $a_\mathrm{max}$ ($\mu$m)&$10^{1,2,3,4,5}$&discr.\\
 turb.~mixing strength $\alpha$&$[10^{-6}, 10^{-1}]$&log.\\\hline \multicolumn{3}{l}{\emph{refinement 1: rounded rim (Eq.~(\ref{eq:rimshape}))}}\\
 transition radius $R_\mathrm{exp}$ (AU)&$[0.5,15]$&lin.\\
 rim width parameter $w$&$[0.05,1]$&lin.\\\hline
 \multicolumn{3}{l}{\emph{refinement 2: large-grain surface density}}\\
 surfdens.~pow.~large $p_{>100\,\mu\mathrm m}$&$[0,2]$&lin.\\\hline
 \end{tabular}
\tablefoot{
\tablefoottext{a}{lin.~= linear; log.~= logarithmic; discr.~= discrete values.}}
\end{table}

\subsection{Data selection}\label{sect:dataselection}
Rather than picking all data to do the fit, we restrict ourselves to a subset for the model runs. As indicated in Sect.~\ref{sect:inspection}, the near-infrared data (PIONIER + NaCo/SAM) are dominated by an overresolved scattering component. The structural information in the data is therefore limited. As we will show below in Sect.~\ref{sect:nearinfrared}, the model visibilities are strongly dominated by our assumptions on scattering (isotropic vs.~anisotropic). Immediately including the near-infrared data would therefore complicate our focus on the disk structure by having to address \emph{simultaneously} this scattering problem. We hence only include the high angular resolution data that are dominated by thermal radiation of the disk, i.e.,~from the mid-infrared to mm-wavelengths. 

Concerning the MIDI data, we averaged the 25-m baseline and the 50-m baseline observations to two correlated flux profiles at an averaged baseline of 26\,m and 48\,m. Also the photometric data were averaged to a single total spectrum (photometric data of 2011 were ruled out in the average due to poor quality). 

Finally, for the SED data, we exclude the VLA 4-cm and 6-cm detections. The photometric points at these wavelengths might be affected by free-free emission, something that will be addressed in Sect.~\ref{sect:excess}.

The total $\chi^2$ associated with the fit is then:
\begin{equation}
 \chi^2=\chi^2_\mathrm{SED}+\chi^2_\mathrm{MIDI,00\,m}+\chi^2_\mathrm{MIDI,25\,m}+\chi^2_\mathrm{MIDI,50\,m}+\chi^2_\mathrm{SMA}+\chi^2_\mathrm{VLA},\label{eq:totchi2}
\end{equation}
where the MIDI-parts of the $\chi^2$ refer to the total spectrum ($00\,$m), averaged 25-m correlated fluxes, and averaged 50-m correlated fluxes.

Below, we describe the process that leads to our final disk model; a graphical representation of the model refinement strategy is shown in Fig.~\ref{fig:refinement}

\begin{figure*}
 \centering
 \includegraphics[width=.49\textwidth,viewport=0 10 550 420,clip]{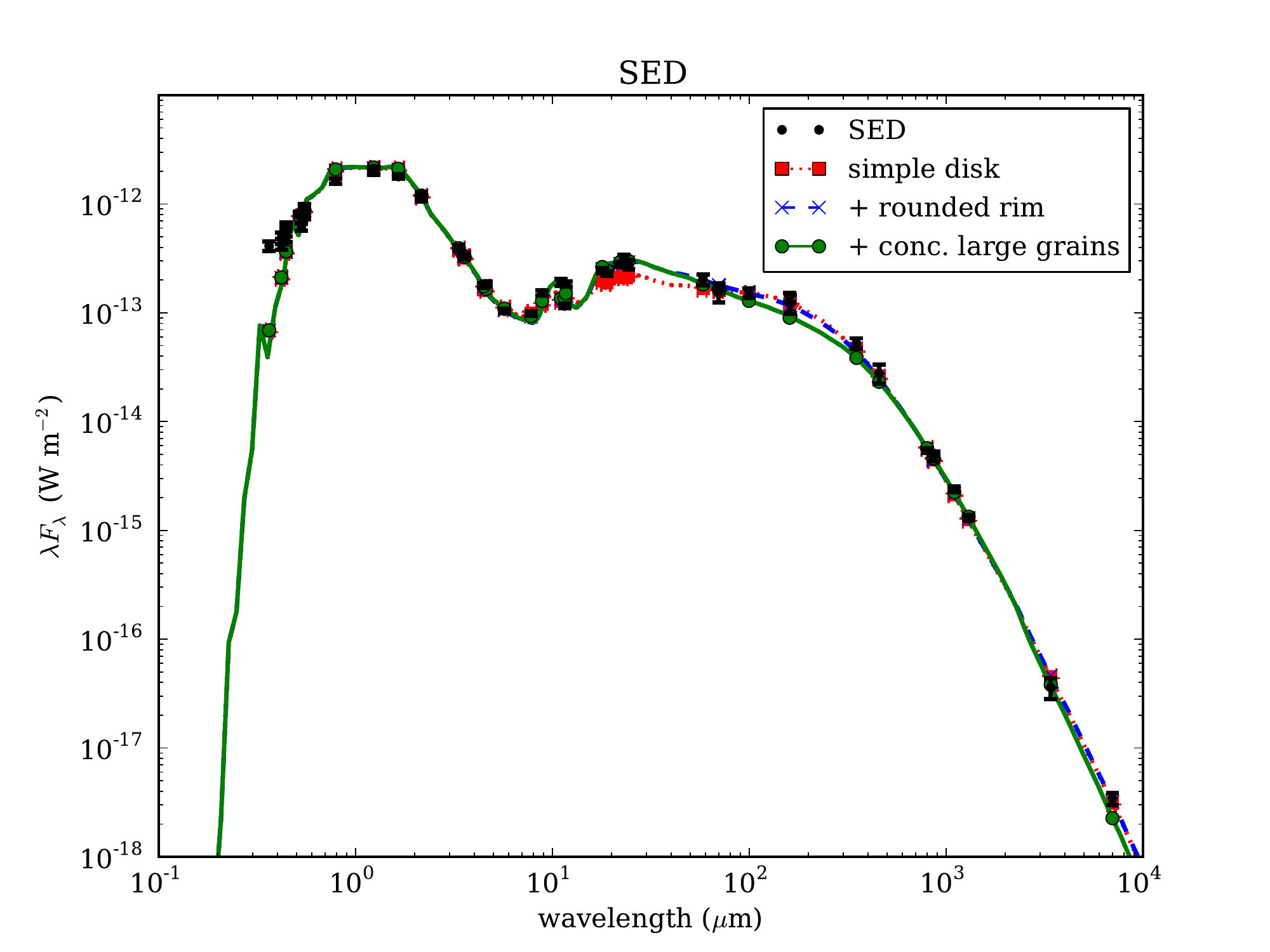}
 \includegraphics[width=.49\textwidth,viewport=0 10 550 420,clip]{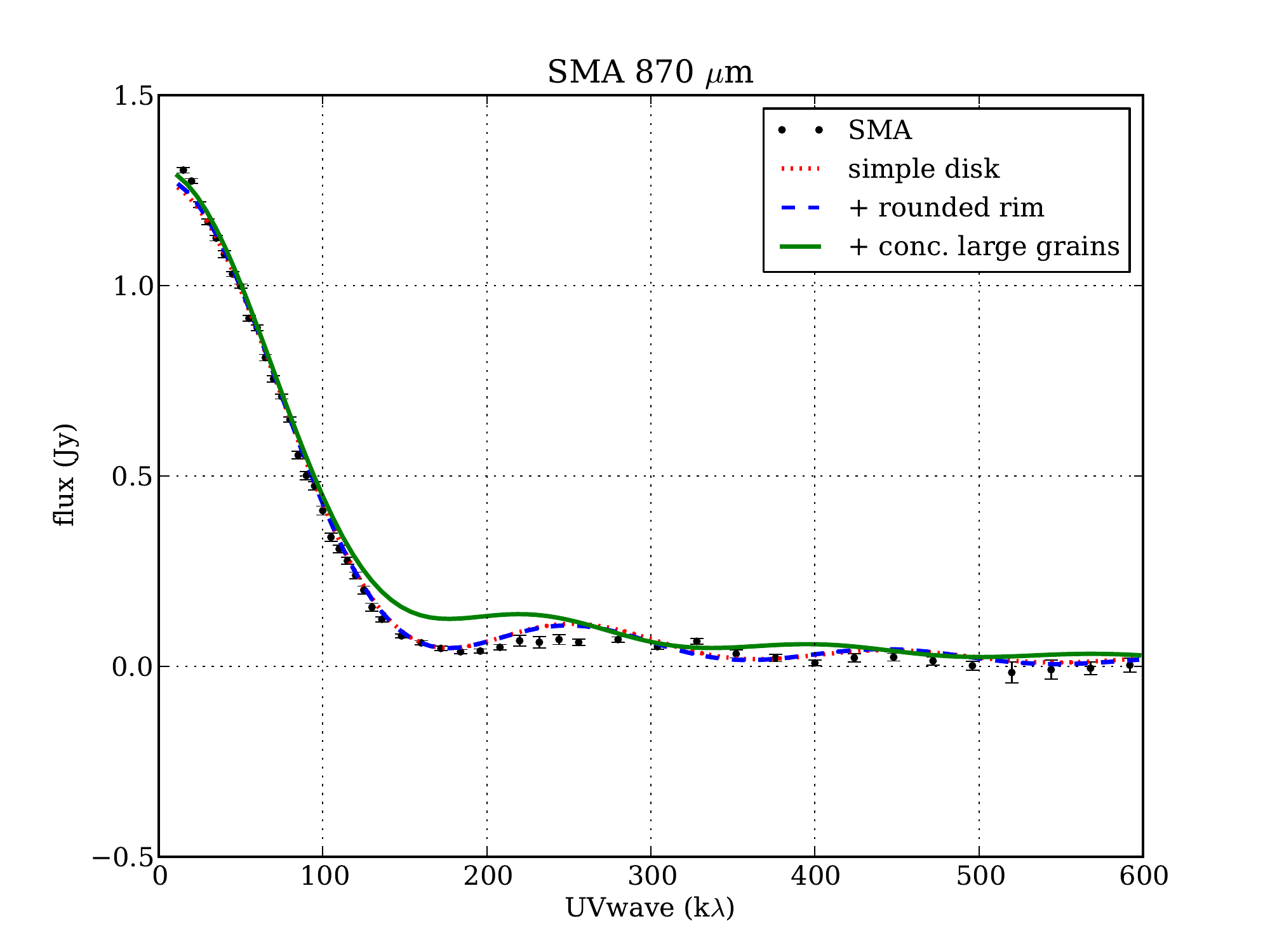}\\
 \includegraphics[width=.49\textwidth,viewport=0 10 550 420,clip]{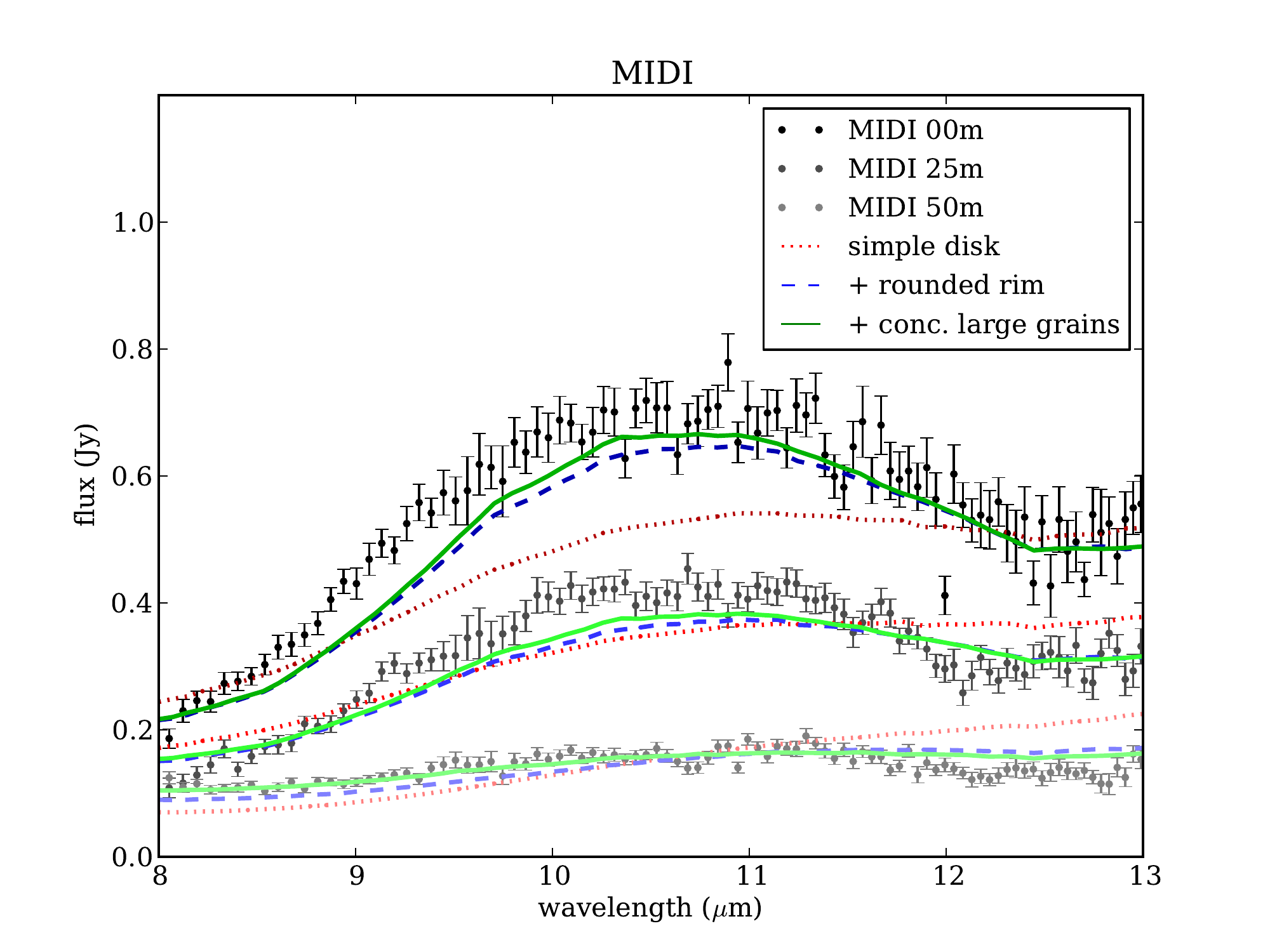}
 \includegraphics[width=.49\textwidth,viewport=0 10 550 420,clip]{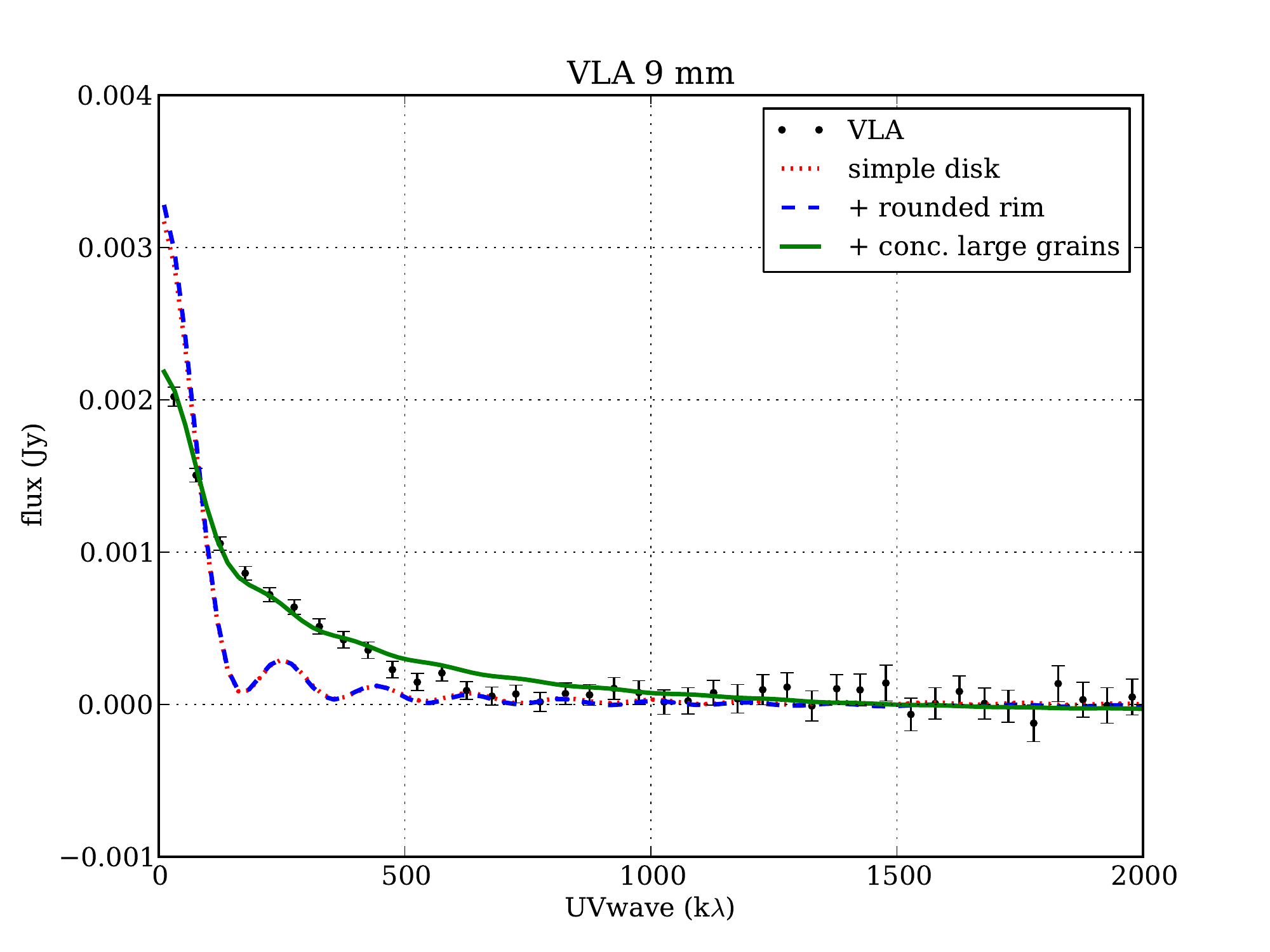}
  \caption{Plots of the best models corresponding to a simple disk (\emph{red, dotted}), a simple disk with a rounded rim (\emph{blue, dashed}), and a simple disk with a rounded rim and a centrally-concentrated large grain population (\emph{green, full}).}\label{fig:singledouble}
\end{figure*}

\subsection{Simple-disk model}\label{sect:simpledisk}
Starting our modeling work from first principles, we begin with a simple radially homogeneous disk model with a vertical inner rim. This model has six free parameters: inner radius $R_\mathrm{in}$, outer radius $R_\mathrm{out}$, dust mass $M_\mathrm{dust}$, surface-density power $p$ (i.e.,~$\Sigma(R)\propto R^{-p}$), maximum grain size $a_\mathrm{max}$, and the turbulence mixing strength $\alpha$. The latter number parametrizes the settling of the dust: the weaker the turbulence, i.e.,~the lower $\alpha$, the stronger the settling of large grains. Since the settling is grain-size dependent, we split the grain population in logarithmic size bins.\footnote{We refer to \citet{2012A&A...539A...9M} for more details on the $\alpha$ parameter and its implementation in \texttt{MCMax}.} The parameter ranges that were tested in the modeling run are shown in Table \ref{table:single}. We note that the ranges cover the physically realistic/interesting values for the specific parameters (e.g.,~testing $R_\mathrm{in}> 1$\,AU is unnecessary, since this would not produce enough flux for simulating the 10-$\mu$m feature). 

\begin{figure*}
 \centering
 \includegraphics[width=\textwidth,viewport=110 0 1310 400,clip]{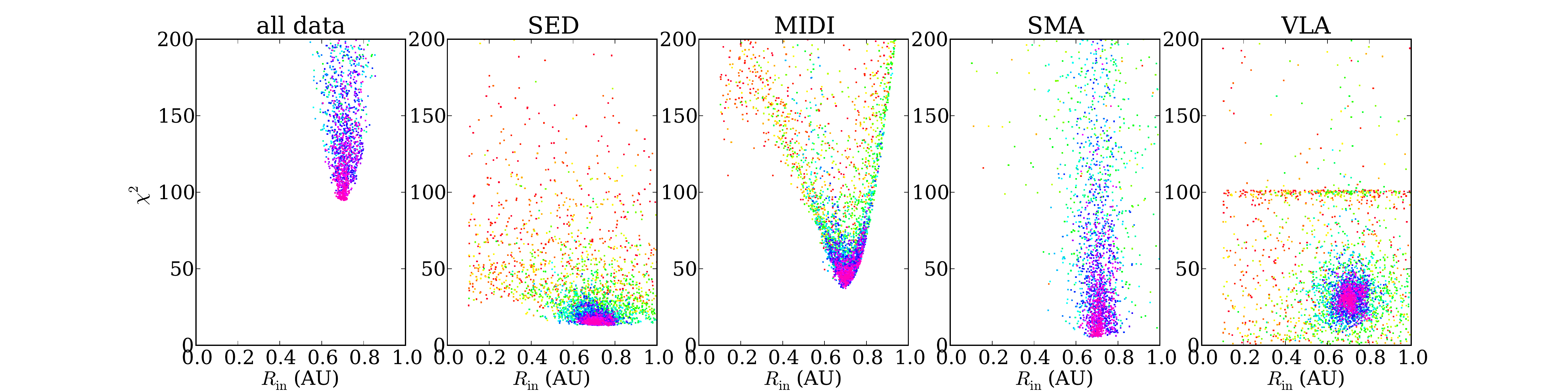}\\\vspace{-0.5cm}
 \includegraphics[width=\textwidth,viewport=110 0 1310 400,clip]{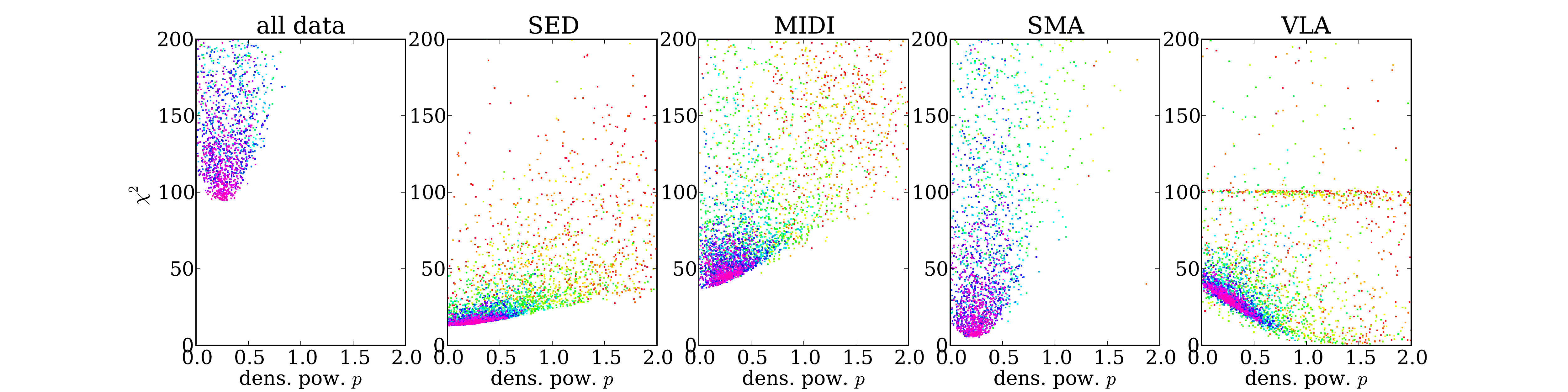}
 \caption{Convergence plots for $R_\mathrm{in}$ (\emph{top}) and the surface density power $p$ (\emph{bottom}), for the simple-disk model. The titles of the individual plots indicate to which data set the shown $\chi^2$ values correspond (e.g.,~for MIDI: $\chi^2_\mathrm{MIDI,00\,m}+\chi^2_\mathrm{MIDI,25\,m}+\chi^2_\mathrm{MIDI,50\,m}$). The colors indicate the model generation to which the points correspond (red = first generations; violet = final generations).} \label{fig:convergence}
\end{figure*}

Fig.~\ref{fig:singledouble} shows the model resulting from calculating 20 generations of 200 models, and the corresponding parameter values are included in Table \ref{table:values}. The fit reaches a reasonable quality for the SED and the SMA data, and the model also reproduces the global flux levels of the MIDI data. However, the visibility profile at VLA wavelengths clearly differs from the actual data (for the non-trivial part of the data): at short baselines, the model shows the oscillatory behavior associated with a sharp disk rim (cf.~the SMA data), whereas the VLA data have a much smoother profile. Nevertheless, the found parameters are interesting, in the sense that they confirm the presence of large grains ($a_\mathrm{max}\ge1\,$mm), a Ratzka-model scale for the inner rim ($\sim0.7\,$AU), and an Andrews-model scale for the outer-disk edge ($\sim60\,$AU).

Table \ref{table:values} contains an estimate of the uncertainty on the different parameters (see Sect.~\ref{sect:parval} and Appendix \ref{appendix:error}). An instructive way to visualize the significance of the different parameters is to make a plot of the $\chi^2$-values of the calculated models vs.~the parameter values corresponding to these models. Using different colors, we can distinguish between the different generations to which the models belong. Going from the first to the last generations, one expects to see a transition from randomly distributed parameter values towards a more concentrated neighborhood, the absolute minimum in the $\chi^2$-space. The plots could therefore be called \emph{convergence plots}, showing the gradual convergence for the different parameters. For unconstrained or degenerate parameters, on the other hand, this behavior will not be seen: other constraints are then needed.

In Fig.~\ref{fig:convergence} we show two different examples of these convergence plots, for $R_\mathrm{in}$ and $p$. The plots show both the total $\chi^2$-values and the $\chi^2$-values of the individual data sets (cf.~Eq.~(\ref{eq:totchi2})). As can be seen in the upper plots, the convergence of $R_\mathrm{in}$ is almost completely determined by the MIDI data, whereas the other data show a more random-like distribution in $\chi^2$-values for different values of $R_\mathrm{in}$.\footnote{It might seem that the SMA data also help to constrain the fit. However, the convergence here is only apparent: essentially all models in the SMA plot belong to later model generations (green to violet color), where $R_\mathrm{in}$ is already well-constrained (by the MIDI data). The SMA data thus do not help constraining $R_\mathrm{in}$.} In the second case, for the surface density power $p$, a rather different behavior is seen. Clearly, the lowest $\chi^2$-values are reached for the lowest $p$-values, as far as the MIDI and the SED data are concerned. By contrast, much lower $\chi^2$-values for the VLA data are reached if larger values ($p>1$) are taken. For the current modeling run, the convergence seems thus to be dominated by the MIDI and SED data, which outweighs the fit to the VLA data.

\subsection{Simple disk with a rounded inner rim}
The MIDI fluxes are mainly determined by the thermal emission of small dust grains at relatively warm temperatures ($\gtrsim100\,$K), thus relatively close to the central star (and close to the disk surface). The convergence to a low value for $p$ in the simple-disk model, dominated by the MIDI data (Fig.~\ref{fig:convergence}), could therefore point to a different density regime in the inner regions of the disk, rather than being representative for the complete disk. Verifying models with steeper surface densities (e.g.,~$p\gtrsim1$) indicates that the inner region of the disks are optically thick in the radial direction up to high scale heights. As a consequence, the sharp, vertical rim strongly intercepts stellar radiation and reradiates it mainly in the radial direction (i.e.,~not in the direction perpendicular to the disk, where the observer is situated). In the low-density case, the radial optical depth at small radii is much smaller, and a significant disk atmosphere is reachable for thermal re-emission along the vertical direction.

As an alternative to a vertical wall, we can introduce a rounded-off wall. The smoother transition from the inner hole to the ``full'' disk avoids the strong geometric effects due to observing the disk pole-on. More into detail, the parametrization of a smooth wall also extends the possibilities for the fitting algorithm to reproduce the MIDI correlated fluxes, and decreases the possible biasing of the fit of the other datasets. 

We utilize the rim structure used in \citet{2013A&A...557A..68M} to model the disk rim of the Herbig Ae/Be star HD\,100546:
\begin{equation}
 \Sigma(R)=\left\{\begin{array}{ll}
\Sigma_\mathrm{exp}\,\bigg(\frac{R}{R_\mathrm{exp}}\bigg)^{-p}\,\exp\left(-\bigg(\frac{1-R/R_\mathrm{exp}}{w}\bigg)^3\right)&\textrm{for }R_\mathrm{in}\le R < R_\mathrm{exp},\\
\Sigma_\mathrm{exp}\,\bigg(\frac{R}{R_\mathrm{exp}}\bigg)^{-p}&\textrm{for }R_\mathrm{exp}\le R \le R_\mathrm{out}.
\end{array}\right.\label{eq:rimshape}
\end{equation}
Here, here $\Sigma_\mathrm{exp}$ is the surface density at the radius $R_\mathrm{exp}$ where the drop in surface density starts, and $w$ is a dimensionless rim width. In other words, we assume that at a certain transition radius, the surface density decreases exponentially towards the inner disk rim. The rim shape has a direct link to hydrodynamical simulations of a transition disk in the presence of a (sub-)stellar companion, to which we go deeper into in Sect.~\ref{sect:rim}. The modification thus introduces two new free parameters: $R_\mathrm{exp}$ and $w$; see Table \ref{table:single}.\footnote{Since the total dust mass $M_\mathrm{dust}$ is a free parameter, $\Sigma_\mathrm{exp}$ is \emph{not} a free parameter.} For simplicity we keep $a_\mathrm{max}$ fixed to 1\,cm, the value found before.

Fig.~\ref{fig:singledouble} shows the resulting best fit after calculating 30 generations of 200 models. A more accurate fit is reached for the MIDI data, and the total $\chi^2$-value lowers from $\chi^2=94.2$ to $\chi^2=67.1$. Yet, some clear difference between the model and the actual data are apparent. As far as the spectrally resolved data are concerned (SED, MIDI), these differences are not necessarily problematic: at least part of the model deviation can be related to the exact dust composition, which is not included in the parameter fitting here and therefore cannot be expected to agree perfectly. However, for the monochromatic visibility data, departures in shape will be mostly related to the exact model for the dust distribution. In the case of the SMA data, a clear departure in shape is seen for projected baselines above $200\,$k$\lambda$. For the VLA data, the departure is even more problematic: as in the simple-disk case, the model has a completely different profile at short baselines.

\begin{figure}
 \centering
\includegraphics[width=.5\textwidth,viewport=20 10 550 400,clip]{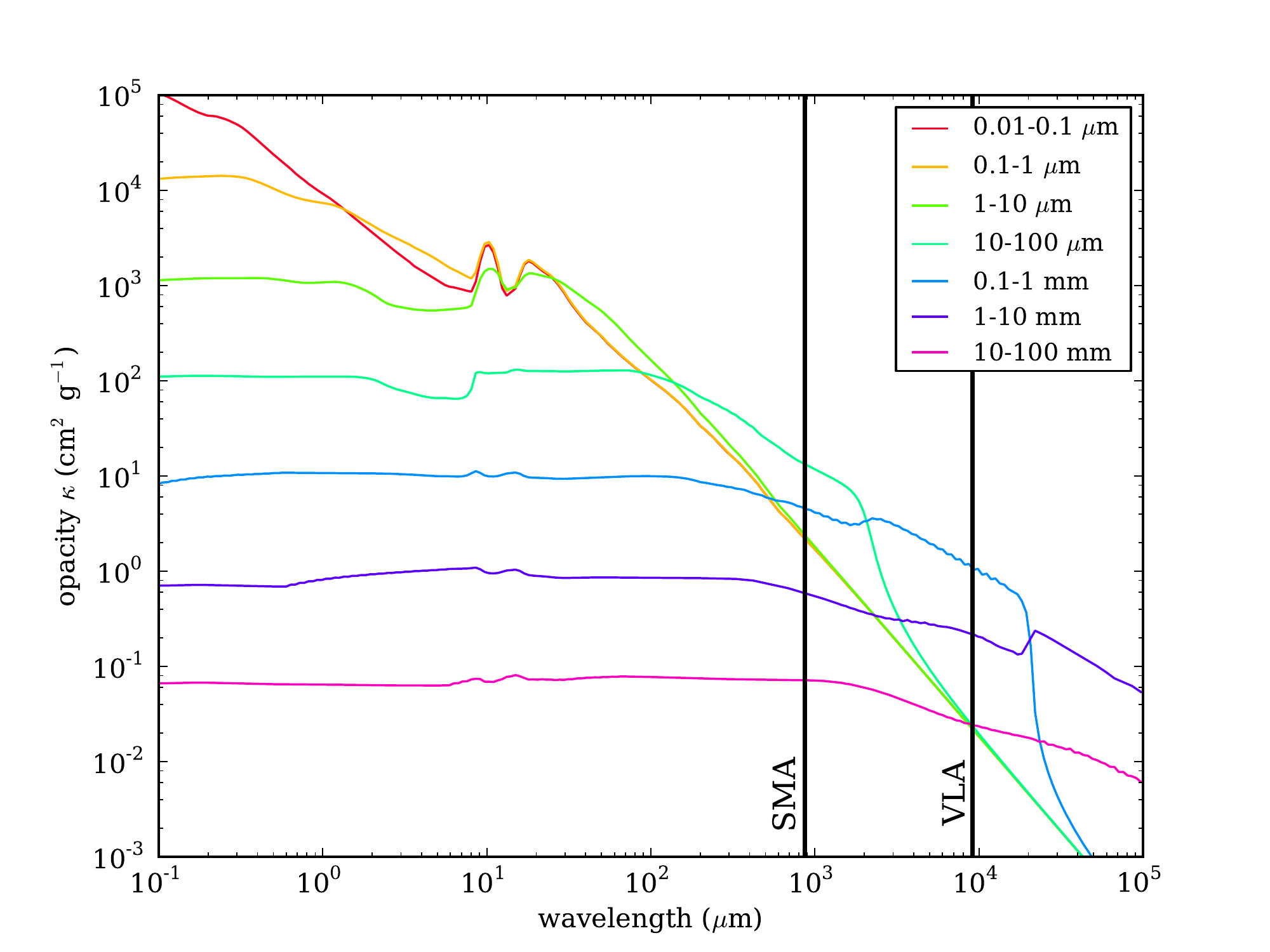}
\caption{Mass absorption coefficients per order of magnitude in grain size, for an $n(a)\propto a^{-3.5}$ size distribution (within each bin). The vertical lines indicate the wavelengths of the SMA and VLA observations, allowing the comparison of the relative opacity contribution for each of the grain size bins.}\label{fig:opacity}
\end{figure}

\begin{table*}
 \centering
 \caption{Parameter values and estimated uncertainties for the best fit of the different models. The values $x_\mathrm{best}$, $\langle x\rangle$, and $\sigma_x$ correspond to the parameter value for the best-fitting model, the parameter value averaged over the marginal probability distribution (i.e.,~the expectation value), and the corresponding 1-$\sigma$ error (i.e.,~$\sqrt{\langle x^2\rangle-\langle x\rangle^2}$). }\label{table:values}{\tiny
\begin{tabular}{l|ccc|ccc|ccc}
\hline\hline
&&\emph{simple disk}&&\multicolumn{3}{|c|}{\emph{+ rounded rim}}&\multicolumn{3}{|c}{\emph{+ concentrated large grains}}\\\hline
&$x_\mathrm{best}$&$\langle x\rangle$&$\sigma_x$&$x_\mathrm{best}$&$\langle x\rangle $&$\sigma_x$&$x_\mathrm{best}$&$\langle x\rangle $&$\sigma_x$\\\hline
\emph{simple disk}&&&&&&&&\\
inner radius $R_\mathrm{in}$ (AU)&$0.70$&$0.68$&$0.04$&$0.54$&$0.52$&$0.09$&$0.32$&$0.35$&$0.10$\\
outer radius $R_\mathrm{out}$ (AU)&$56.1$&$56.8$&$1.7$&$56.3$&$57.1$&$1.6$&$61.7$&$60.8$&$1.9$\\
dust mass $M_\mathrm{dust}$ (M$_\odot$)\tablefootmark{a}& $1.8\times10^{-4}$&$1.6\times10^{-4}$&${}^{+0.3}_{-0.2}\times10^{-4}$ &$1.9\times10^{-4}$&$1.9\times10^{-4}$&$0.3\times10^{-4}$&&&\\
$\rightarrow$ $M_{\mathrm{dust},<100\,\mu\mathrm m}$& &&& &&& $4.6\times10^{-5}$&$5.2\times10^{-5}$&${}^{+2.3}_{-1.6}\times10^{-5}$\\
$\rightarrow$ $M_{\mathrm{dust},>100\,\mu\mathrm m}$& &&& &&& $5.0\times10^{-5}$&$4.4\times10^{-5}$&${}^{+1.5}_{-0.8}\times10^{-5}$\\
surfdens.~pow.~$p$&$0.27$&$0.32$&$0.09$&$0.31$&$0.48$&$0.18$&$0.5$&$0.6$&$0.2$\\
max.~grain size $a_\mathrm{max}$ ($\mu$m)&$10^4$&$\geq10^3$&-& $10^4$&\multicolumn{2}{c|}{-} &$10^3$&\multicolumn{2}{c}{-} \\
turb.~mix strength $\alpha$&$0.8\times10^{-5}$&$1.9\times10^{-5}$& ${}^{+2.4}_{-1.0}\times10^{-5}$&$0.5\times10^{-5}$&$0.9\times10^{-5}$&${}^{+0.7}_{-0.4}\times10^{-5}$&$1.0\times10^{-5}$&\multicolumn{2}{c}{-} \\
\hline \emph{+ rounded rim}&&&&&&&&\\
 transition radius $R_\mathrm{exp}$ (AU)&&/&&$3.1$&$3.3$&$0.6$&$3.1$&$3.1$&$0.5$\\
 rim width parameter $w$&&&&$0.52$&$0.49$&$0.10$&$0.45$&$0.45$&$0.06$\\\hline
\emph{+ concentrated large grains}&&&&&&&&\\
surfdens.~pow.~large $p_{>100\,\mu\mathrm m}$&&/&&&/&&$1.3$&$1.4$&$0.2$\\\hline
\end{tabular}
\tablefoot{
\tablefoottext{a}{$M_{\mathrm{dust},<100\,\mu\mathrm m}$ and $M_{\mathrm{dust},>100\,\mu\mathrm m}$ are the masses of the dust grains smaller and larger than $100\,\mu$m, respectively. Only for the final model run (with a separate distribution of the $>100$-$\mu$m grains) the total dust mass parameter $M_\mathrm{dust}$ was decoupled into these two fit parameters. }}
}
\end{table*}

\subsection{\textrm{Intermezzo}: sub-mm vs.~mm-visibilities}
The current model geometry has difficulties in simultaneously accommodating the sub-mm and the mm-data. The convergence to the final model is currently mainly determined by the mutual weight of the individual data sets, and clearly the SMA data have more weight in the fit than the VLA data.

The visibility profiles of the SMA and VLA data in Fig.~\ref{fig:singledouble} indicate that the disk looks very different in the two wavelength regimes. Since the disk is optically thin at both wavelengths, the dominant opacity sources must be distributed in a different way throughout the disk (i.e., the difference in visibilities is not an optical-depth effect). The steep visibility decrease at sub-mm wavelengths with respect to the smooth decrease at mm-wavelengths is evidence of a more compact distribution of the region emitting in the millimeter. Also, the visibility oscillations in the SMA data are not apparent in the VLA data, indicating that we do not have sharp edges within the spatial resolution range of the VLA observations (at least, at the mm-wavelengths).

One way to make a smoothly decreasing visibility profile with centrally concentrated emission is to make a centrally concentrated dust distribution that rapidly decreases outwards. Since the surface density is steeply decreasing with radius, the outer edge of the disk will not be in high contrast with the region devoid of dust, avoiding the strong oscillatory behavior (``ringing'') in the visibility curve. However, the above distribution should only concern the mm-emission from the dust: for the sub-mm data, we will still need a dust distribution with a sharp outer edge.

\subsection{Compact distribution of the largest grains}
From our above reasoning, it seems essential to distinguish between two dust populations: one determining the disk at sub-mm wavelengths vs.~one that sets the brightness distribution at mm-wavelengths. For the implementation of the dust settling, we have split up the full grain distribution in bins of an order of magnitude in grain size ($0.01-0.1\,\mu$m, $0.1-1.0\,\mu$m, etc.; see Sect.~\ref{sect:simpledisk}). We show a plot of the opacity curves $\kappa_\lambda$ of each of the size bins in Fig.~\ref{fig:opacity}. It is clear that the grain size $\sim100\,\mu$m determines a transition between size bins that have $\kappa_\mathrm{SMA}\gg\kappa_\mathrm{VLA}$ and $\kappa_\mathrm{SMA}\sim\kappa_\mathrm{VLA}$. Therefore, most of the disk appearance at sub-mm wavelengths will be determined by the sub-100-$\mu$m dust. This allows us to deliberate our large-grain population, and freely distribute it throughout the disk, without considerably disturbing the visibilities at sub-mm wavelengths.

We propose to implement the redistribution of the $>100$-$\mu$m dust by introducing a new parameter, the surface density power $p_{>100\,\mu\mathrm m}$ for these large grains. We assume the \emph{same} surface-density profile as in Eq.~(\ref{eq:rimshape}) for these large grains, only with a different power $p_{>100\,\mu\mathrm m}$ (i.e.,~$p_{>100\,\mu\mathrm m}\ne p$). 

The final model run deserves the following comments:
\begin{itemize}
 \item The above models clearly showed the SMA fit to outweigh the fit of the VLA data. This is an intrinsic issue due to the relative level of accuracy of the former data set with respect to the latter. In order not to favor one data set over the other, we readjust the relative weights of the two data sets by decreasing and increasing them by a factor of 2, respectively (i.e.,~$\chi^2_\mathrm{SMA}\rightarrow \frac12\chi^2_\mathrm{SMA}$; $\chi^2_\mathrm{VLA}\rightarrow 2\,\chi^2_\mathrm{VLA}$). Inspection of the fits \emph{a posteriori} shows that this factor leads to a satisfactory reproduction of all data sets (at least by eye).
 \item We found it necessary to fix two parameters to get a valid fit. First, we fixed $a_\mathrm{max}$ to 1\,mm. As can be inferred from the previous model fits (the shown models in Fig.~\ref{fig:singledouble}), the VLA-visibilities at short baselines are overpredicted for $a_\mathrm{max}=1\,$cm. Taking $a_\mathrm{max}=1\,$mm lowers the mm-flux, which allows the model to reach valid visibility levels. The second parameter we fix is the settling parameter $\alpha$. The strongest diagnostic for $\alpha$ in our data set is the SED shape in the range $\lambda=100-500\,\mu$m: this wavelength range corresponds to the thermal emission of the outer flaring disk, and it is exactly $\alpha$ which determines the amount of flaring in the different dust populations. However, test model runs show that subtle effects at longer wavelengths (with now two equally contributing visibility data sets) outweigh the determination of $\alpha$ from the few SED points in the range $\lambda=100-500\,\mu$m. We therefore fix $\alpha=10^{-5}$, the relatively well constrained value found before.
 \item We limit the model fitting algorithm to a compact parameter space around the best parameters of the previous models. The only parameters we give a large freedom are the surface density powers $p$ and $p_{>100\,\mu\mathrm{m}}$, and the dust masses $M_{\mathrm{dust},<100\,\mu\mathrm{m}}$ and $M_{\mathrm{dust},>100\,\mu\mathrm{m}}$. We note that we indeed split up the dust masses in the small and large grains, since the different spatial distribution for the two grains populations does not necessarily mean that their relative abundance should be the same as for a radially homogeneous disk.
\end{itemize}

The best fit model after calculating 30 generations of models is shown in Fig.~\ref{fig:singledouble}. The assumption of a differently distributed large-grain population somewhat degrades the fit of the SMA data, which confirms that the sub-mm opacity of the largest grains contributes significantly. In other words, the idea of decoupling the two grain populations to fit the two visibility curves independently was (obviously) idealized. Still, we manage to strongly improve the fit quality of the VLA data as compared to the radially homogeneous models, reproducing the data both qualitatively and quantitatively. The total $\chi^2$-value lowers to $\chi^2=51.4$ (compared to $\chi^2=67.1$ for the previous best model).

\section{Discussion}\label{sect:discussion}
\subsection{Parameter validity range}\label{sect:parval}
The outcome of a modeling run is a set of parameters, corresponding to the best-fitting model. Getting an idea on \emph{how well} the model fits the data, i.e.,~determining the uncertainty on the determined parameters, is a non-trivial problem. 

Different techniques exist for determining the validity range of fit parameters (see, e.g.,~\citet{2010arXiv1009.2755A} for a concise overview). Several of the techniques are based on a relatively extended sampling of the parameter space (e.g.,~Markov Chain Monte Carlo methods, parameter grid methods), or make use of refitting the slightly modified data set (e.g.,~resampling, bootstrapping techniques). In the case of highly dimensional parameter spaces, with computationally expensive model calculations, most of the techniques become highly demanding in terms of computation time/power. 

The reason for using a genetic fitting algorithm for our modeling work was exactly to limit the computational effort needed. We therefore also have interest in finding a method that gives error estimates in a reasonably simple way. Several methods have been used to find validity ranges for parameters when using a genetic algorithm for fitting. \citet{2009A&A...501.1259C} use a data-resampling technique for generating artificial data sets, which are then refitted. \citet{2013A&A...550A..74D} also make use of additional fits, but each time on the same data set (here, the uncertainty is reduced to the reproducibility of the best model).\footnote{In essence, this method is therefore not a real ``error'' estimation: in the ideal case where the algorithm each time traces back the same best model, this would imply that the parameter uncertainty is 0, which is obviously unrealistic.} For both methods, new fits are thus required, multiplying the computation time by the number of redone fits. A different approach is to specify an \emph{ad hoc} parameter $\Delta\chi^2$, defining a range of calculated models and hence parameters that are still ``valid'' \citep{2011MNRAS.415.2953J}. 

We make use of a method based on the collection of models that is calculated to get to the best model. The method is explained in Appendix \ref{appendix:error}. In essence, the parameters of the calculated models are resampled on a regular grid, which allows us to use a Bayesian inference method as in, e.g.,~\citet{2008A&A...489..633P} or \citet{2012A&A...539A..17L}.

\subsection{Parameter values of different models}
Table \ref{table:values} allows us to compare the values of the parameters shared by the subsequent model geometries, and compare with previously proposed disk models. 

The simple-disk model with a vertical rim has an inner-disk geometry that is well constrained, with an inner radius of $0.7\,$AU. The scale is exactly what was found by \citet{2007A&A...471..173R}, and also confirms our simple-ring fit in Sect.~\ref{sect:ringfit}. Our radiative transfer modeling shows that a vertical rim is not properly explaining the full 10-$\mu$m emission profile in the MIDI correlated fluxes, something that is improved upon by introducing a dust-depleted inner region (i.e., the rounded rim). This moves in the inner radius to a less constrained value of $0.3-0.5\,$AU. The starting point of the ``rounding-off'' of the inner rim lies around $\sim3\,$AU, as can be inferred from both model geometries that include the rounded rim. We note that $R_\mathrm{exp}$ does \emph{not} correspond to the radius of the maximum surface density. The latter lies closer to the central star, roughly around $2.5\,$AU for the disk models with $R_\mathrm{exp}\sim3\,$AU. The resulting surface-density profile, which is then smoothly decreasing inwards of this radius, is thus rather different than the profile with a sharp transition at 4\,AU, typically used to model the disk (e.g.,~\citealt{2010A&A...518L.125T}; \citealt{2012ApJ...744..162A}; \citealt{2012ApJ...750..119A}; \citealt{2013ApJ...766...82Z}).

For $R_\mathrm{out}$, we find relatively well constrained values around 60\,AU for the three tested model geometries. These values confirm the value of 60\,AU found by \citet{2012ApJ...744..162A}. As was already mentioned by these authors, the value differs from the outer radius of the disk as detected in scattered light and CO ($>200\,$AU). We go deeper into this difference in Sect.~\ref{sect:outerradius}, where we also describe the constraints coming from our near-infrared data.

A final parameter we consider here is the total dust mass $M_\mathrm{dust}$. For the first two model geometries, we find a dust mass around $2\times10^{-4}\,$M$_\odot$; for the final geometry, the total dust mass is around $1\times10^{-4}\,$M$_\odot$. The difference between the two is due to the maximum grain size $a_\mathrm{max}$ in the models (1\,cm vs.~1\,mm): the higher $a_\mathrm{max}$, the lower the mass-averaged absorption coefficients (since small grains dominate the absorption/emission at short wavelengths). Comparing the mass of our disk model with the masses of the reimplemented models in Sect.~\ref{sect:reimplementing} shows that our estimate is relatively low. The difference in mass estimates is strongly related to the different opacities assumed in the different models (Fig.~\ref{fig:modelopacities}). Since the disk is optically thin at long wavelengths, the emission at long wavelengths is proportional to the mass and the opacity at these wavelengths. For a given millimeter flux, a higher assumed opacity will therefore lead to a lower dust mass. This is exactly what we see here: our model has high dust opacities at millimeter wavelengths, and this results in a relatively low dust mass. Any comparison of model dust masses therefore translates into the inherent uncertainties on dust properties. 

\begin{figure}
\includegraphics[width=.5\textwidth,viewport=120 70 480 380,clip]{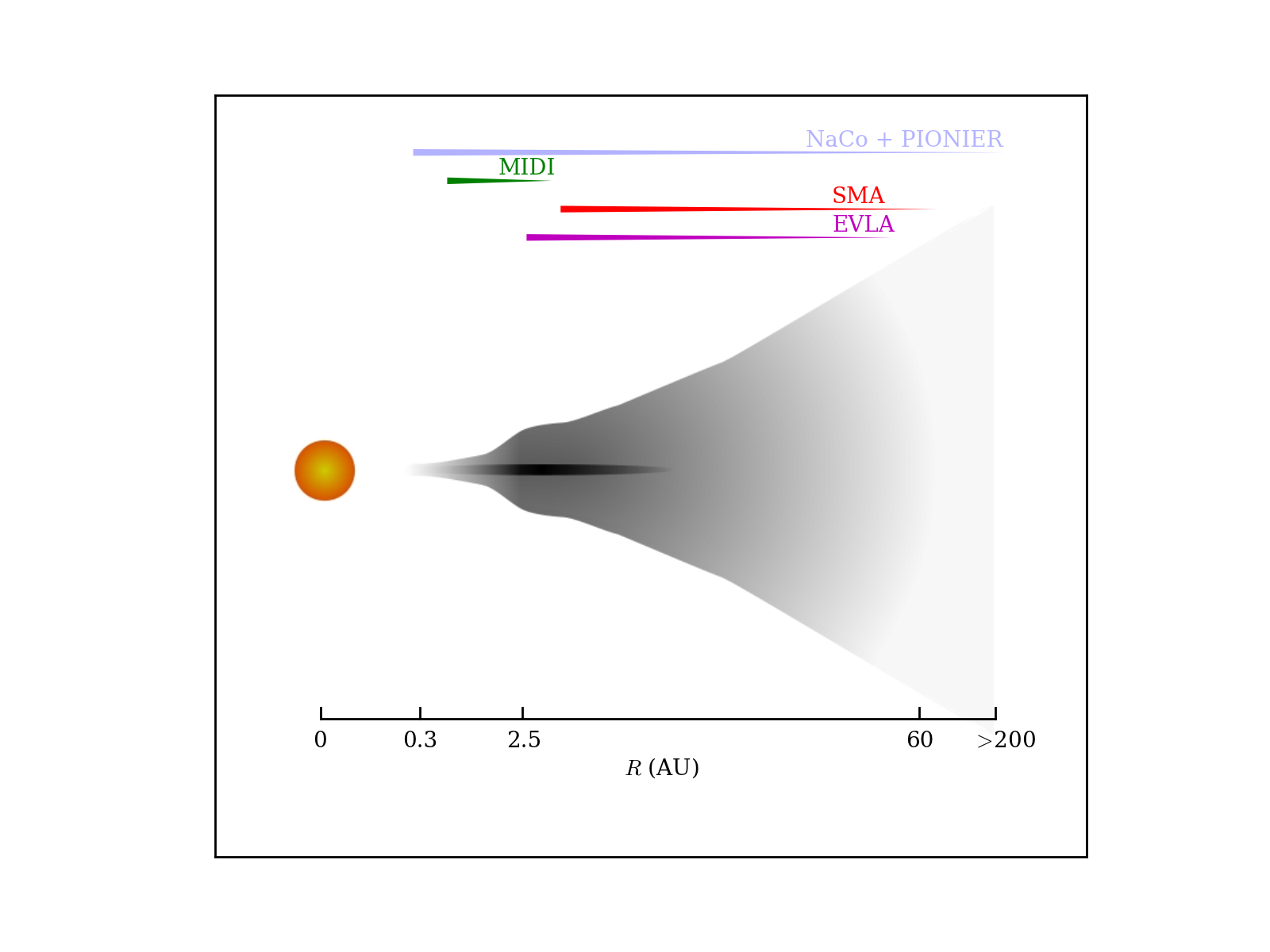}
 \caption{Current image of the TW\,Hya transition disk. In black, we represent the centrally concentrated large-grain population, which is also strongly vertically settled. The light-gray outer-disk region between 60\,AU and $>200$\,AU is not constrained by our radiative transfer model, but is discussed in Sect.~\ref{sect:outerradius}. We also indicate the regions constrained by the different data sets. The near-infrared data provide only weak constraints (see Sect.~\ref{sect:nearinfrared}).}\label{fig:final_drawing}
\end{figure}

A graphical representation of our newly proposed disk model is shown in Fig.~\ref{fig:final_drawing}.

\subsection{Rim shape}\label{sect:rim}
Studying the geometry of the inner region of protoplanetary disks is challenging, in particular concerning the shape of the inner disk rim \citep{2010ARA&A..48..205D}. Initial radiative-transfer models with vertical inner rims have near- and mid-infrared excesses that are strongly inclination dependent \citep{2001ApJ...560..957D}. Later models for the inner rim self-consistently led to a rounded rim structure, a consequence of the temperature dependence of dust condensation on gas density \citep{2005A&A...438..899I}. 

In the case of transition disks, the radius of the disk rim is (much) larger than the typical condensation radius of the dust grains and is hence \emph{not} determined by the condensation temperature of dust. A possible hypothesis is that the inner rim of the transition disk is shaped by the presence of a (sub-)stellar companion inside the gap. Recently, \citet{2013A&A...557A..68M} made a detailed study of MIDI observations of the (pre-)transition disk object HD\,100546. They showed that the observations agree with a disk with the surface-density profile in Eq.~(\ref{eq:rimshape}). Hydrodynamical simulations of the inner disk region then show that the surface density can be reproduced by assuming a $\sim60\,$M$_\mathrm J$ companion within the disk gap.

Our correlated-flux profiles do not contain the clear diagnostics as present in the HD\,100546 data in \citet{2013A&A...557A..68M}, but we do show that the smooth surface-density profile leads to a better reproduction of the correlated fluxes than assuming a sharp transition. This indicates that a companion \emph{might} have shaped this inner rim.  Properties of a possible companion could be derived using similar hydrodynamical simulations of the inner-disk region, something that is beyond the scope of this work. A planetary body within the central $\sim0.1\,$AU could agree with the proposed companion to TW\,Hya at $0.04\,$AU \citep{2008Natur.451...38S}, which existence is yet under debate \citep{2008A&A...489L...9H}.

Although a companion hypothesis is attractive, other processes might contribute to (or even dominate) the creation of the inner cavity. There is strong evidence that photoevaporation plays an important role in the dispersal of the inner disk of TW\,Hya \citep{2011ApJ...736...13P}. Moreover, \citet{2013arXiv1312.4293B} make clear that the gaps carved out in disks by companions of a Jupiter mass or less will be narrow annuli, and not cavity-like. Full dynamical simulations of the inner disk region, taking previous results on photoevaporation and our constraints on the dust into account, will be an important next step.

\subsection{A compact distribution in the largest grains}
Fig.~\ref{fig:singledouble} showed that the visibility profiles at sub-mm and mm-wavelengths differ significantly. Differences in visibility profiles at different mm-wavelengths are seen for other T Tauri stars, and led to a common approach of modeling the data at different wavelengths separately (e.g.,~\citealt{2010ApJ...714.1746I}). The differences in derived surface-density profiles can be interpreted in terms of a radially varying $\beta$, the opacity slope at millimeter wavelengths ($\kappa_\lambda\propto\lambda^{-\beta}$). This change in millimeter opacities is then seen as evidence that grain sizes are decreasing with distance from the central star \citep{2011A&A...529A.105G,2012ApJ...760L..17P}, a conclusion that is also reached for younger (Class 0) objects \citep{2009ApJ...696..841K}.

The aim of this work is to find a radiative transfer model that \emph{directly} incorporates the dust emission at different wavelengths. Making models with a radially homogeneous grain composition, i.e.,~with the same relative abundances for the grain-size bins throughout the disk, clearly showed to be unsuccessful. Instead, the visibility profiles inspired us to propose to \emph{decouple} the largest grains from the smaller-grain population. This, in turn, resulted in a model with two different surface density regimes. For the grains with sizes below $100\,\mu$m, the expectation value for the surface-density power is $p=0.6\pm0.2$ (the best-fit model has $p=0.5$). This is close to the value 0.75 found for the Andrews model, see Table \ref{table:reimplementation}, based on the same SMA data set. For the larger grains ($a>100\,\mu$m), the surface density power in the best model is two times higher.

The spatial and size distribution of dust grains in protoplanetary disks is determined by the processes of dust growth, fragmentation, and transport in viscously evolving gas disks. \citet{2012A&A...539A.148B} present a simple model for the dust evolution in disks that agrees with high-level simulations of the incorporated processes \citep{2010A&A...513A..79B}. Two limiting cases are determined: the fragmentation-limited distribution, where the dust particles are in a steady state in which coagulation and fragmentation balance, and the drift-dominated distribution, where dust particles are drifting away more rapidly than being replenished by growth. The dominant regime in a certain part of the disk depends on parameters like the turbulence level, the velocity at which grains fragment, and the initial gas-to-dust ratio. 

\citet{2012A&A...539A.148B} show that the \emph{drift-dominated} disks have a dust surface density proportional to $R^{-0.75}$, for gas-disk profiles with $\Sigma_\mathrm{gas}\propto R^{-1}$, and point out that the Andrews model agrees with this surface-density profile. As was already indicated, our surface density for the $<100$-$\mu$m grains agrees with this profile, except that our assumption of a constant dust-to-gas ratio leads to a gas surface density of the form
\begin{equation}
 \Sigma_\mathrm{gas}(R)\propto A\,R^{-0.6}+B\,R^{-1.4},
\end{equation}
where $A$ and $B$ are constants. However, the sum of the shallow and the steep density profile leads to a surface density that differs by less then $10\,\%$ from a ``regular'' $R^{-1}$ profile, for almost the complete disk. This implies that the approximation for this simple drift-dominated disk model is valid. Concerning now the grain population larger that $100\,\mu$m, we have a surface-density profile that is considerably different from the drift-dominated case. Interestingly, the found expectation value $p_{>100\,\mu\mathrm m}=1.4\pm0.2$ (and $p_{>100\,\mu\mathrm m}=1.3$ for the best-fitting model) is very close to surface-density power 1.5 corresponding to the \emph{fragmentation-dominated} distribution \citep{2012A&A...539A.148B}. It seems therefore that the different traced surface densities can be related to physically different regimes within the disk. 

For typical simulation results in \citet{2012A&A...539A.148B}, disks with an age of a few Myr have maximum particle sizes that are drift-dominated in the outer regions and fragmentation-dominated in the inner regions. In addition, the models also show that the largest grains are found in the inner region of the disk. The results we get from our radiative-transfer modeling seem to confirm this picture. For the largest grains, thus probing the inner disk region, we indeed find a surface density that has a fragmentation-dominated character; for the smaller grains, probing also further-out regions in the disk, we seem to confirm a drift-dominated surface density.

\subsection{Linking the dust emission to gas properties}
Our analysis has implicitly substituted ``modeling the disk structure'' by ``modeling the dust distribution''. The contribution of the gas to the disk structure is only incorporated in terms of the vertical hydrostatic equilibrium and the vertical settling of the dust grains. In order to investigate the gas and dust distribution simultaneously, gas-emission diagnostics would be required, preferably spatially resolved (see, e.g.,~\citealt{2012ApJ...744..162A} for the analysis of the outer disk in CO). Still, a few indirect links can be exploited to couple the disk model to the intrinsic gas properties.

By its nature, the turbulent mixing strength parameter $\alpha$ couples the gaseous structure of the disk to the dust distribution. \citet{2012A&A...539A...9M} model median SEDs of Herbig stars, T Tauri stars, and brown dwarfs, and find $\alpha=10^{-4}$ to be a representative value across this range of stellar masses (their analysis is based on the same radiative-transfer code as the one used here). The here inferred value of $\alpha\sim10^{-5}$ seems to indicate that TW\,Hya has a more-settled disk than average. \citet{2011ApJ...727...85H} model the spatially resolved CO mm-emission of TW\,Hya, and estimate that $\alpha\sim10^{-3}-10^{-2}$, in line with accretion-based estimates of $\alpha$ for other protoplanetary disks ($\alpha\sim10^{-2}$: \citealt{1998ApJ...495..385H}). The gas-based (i.e.,~CO emission, accretion) estimates of $\alpha$ clearly indicate a higher value than the $\alpha$-value found here. Although more work on reconciling the two approaches is needed, this apparent inconsistency might simply indicate that the assumption of a single, constant $\alpha$ throughout the disk is most likely invalid (see, e.g.,~dead zones). Therefore, different estimates might be dominated by specific disk regions, which does not necessarily gives the same values.

A second diagnostic for the gas is offered by the temperature structure of the disk, a direct output of the radiative-transfer model. ALMA observations of TW\,Hya have recently led to the discovery of the CO ice line at $\sim29\,$AU, by inferring the inner-edge location of N$_2$H$^+$ in the midplane \citep{Qi18072013}. These observations provide an independent measure for the midplane temperature. Our radiative transfer model has a midplane temperature of 14\,K at $29\,$AU. This value is reasonably close to the temperature of $\sim17\,$K inferred for the N$_2$H$^+$ inner-edge location, agreeing with expected CO sublimation temperatures \citep{Qi18072013}. This indicates that our purely dust-based model leads to a temperature structure that might be in reasonable agreement with the temperature structures inferred using gas-diagnostics, at least for regions where gas and dust temperature are similar.

\subsection{Contribution of excess radio continuum emission}\label{sect:excess}
The data analysis presented here is restricted to $\lambda<1\,$cm, since emission at cm-wavelengths is possibly affected by the presence of an ionized wind (Sect.~\ref{sect:sed}). Here, we shortly discuss the nature of the radio continuum excess, and justify our model assumption of pure dust emission at sub-cm wavelengths, in particular for the resolved 9-mm VLA data.

\citet{2012ApJ...751L..42P} analyze the contribution of free-free emission in the long-wavelength SED of TW\,Hya. They conclude that the 7-mm emission agrees with pure dust emission, whereas the 3.5-cm emission \citep{2005ApJ...626L.109W} clearly shows an excess. The new 4.1-cm and 6.3-cm VLA detections of TW\,Hya allow us to repeat the \citet{2012ApJ...751L..42P} analysis, and fit the actual contribution of the cm-excess.\footnote{The analysis of \citet{2012ApJ...751L..42P} is based on an upper limit for the 6-cm emission, which implies that the spectral index of the cm-excess was only partly constrained.} Following their approach and other arguments, the following conclusions can be made on the long-wavelength SED of TW\,Hya (Appendix \ref{appendix:longSED}):
\begin{itemize}
 \item the contribution of a radio excess at $\lambda=9\,$mm (i.e., the wavelength of the modeled VLA observations) is low, perhaps on the level of 5\,$\%$;
 \item the potential radio-excess source cannot be point-like;
 \item even at wavelengths as long as 4\,cm, the dust emission seems to be detected, confirming the earlier suggestion by \citet{2005ApJ...626L.109W} that for TW\,Hya, dust emission is important even at wavelengths of 3.5\,cm.
\end{itemize}
The non-point-like nature of the excess implies that any attempt to model this excess would require including a parametrized geometry of the responsible region. This is a non-trivial issue, and the data quality is likely not good enough to assess this problem. Taking the uncertainty level of the data and the arguably small excess contribution into account, we can therefore justify our modeling assumption of pure dust emission at 9\,mm, i.e.,~the use of a pure radiative-transfer model.

\begin{figure*}
 \centering
 \includegraphics[width=.49\textwidth,viewport=0 10 550 440,clip]{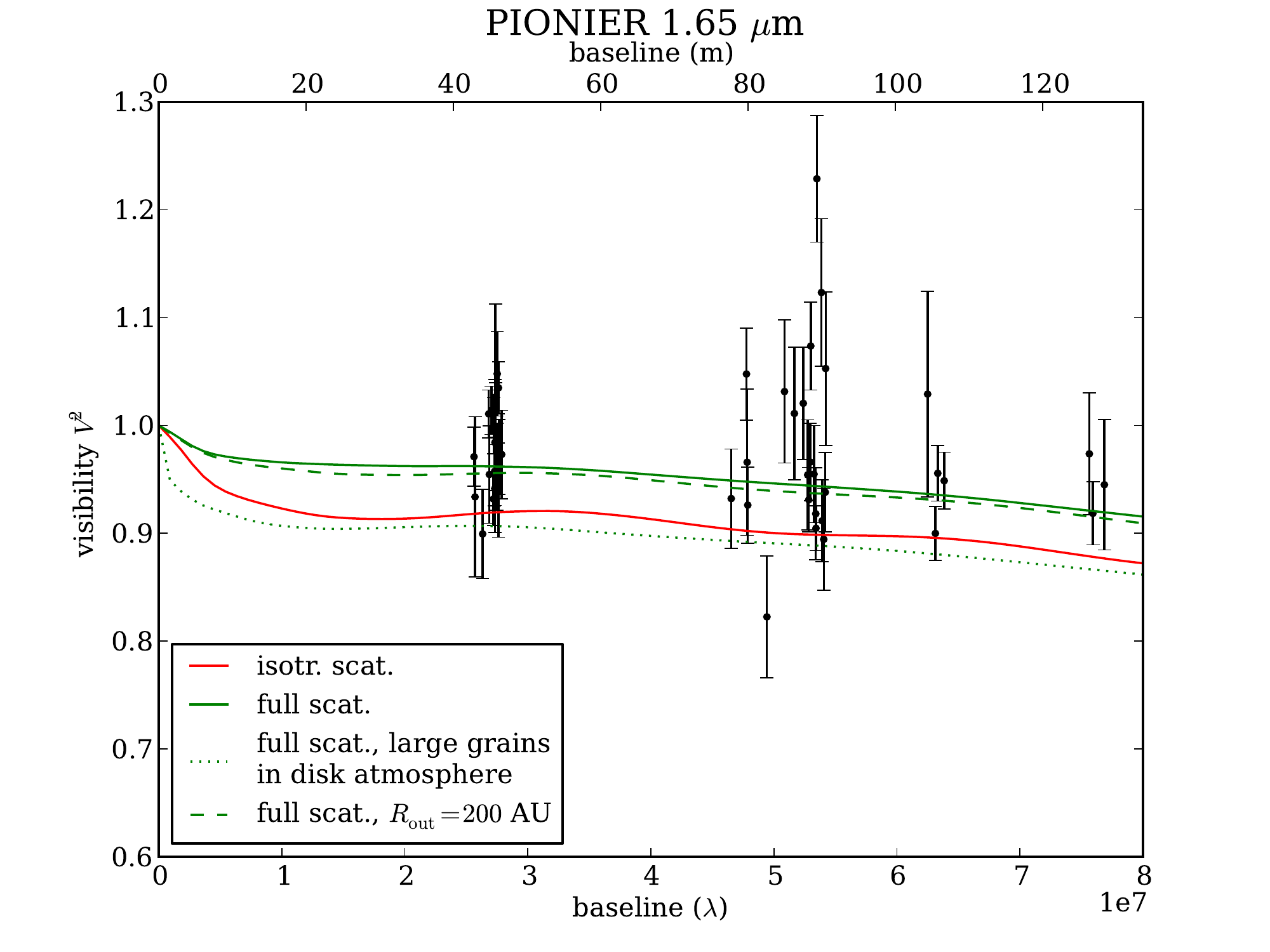}
 \includegraphics[width=.49\textwidth,viewport=0 10 550 440,clip]{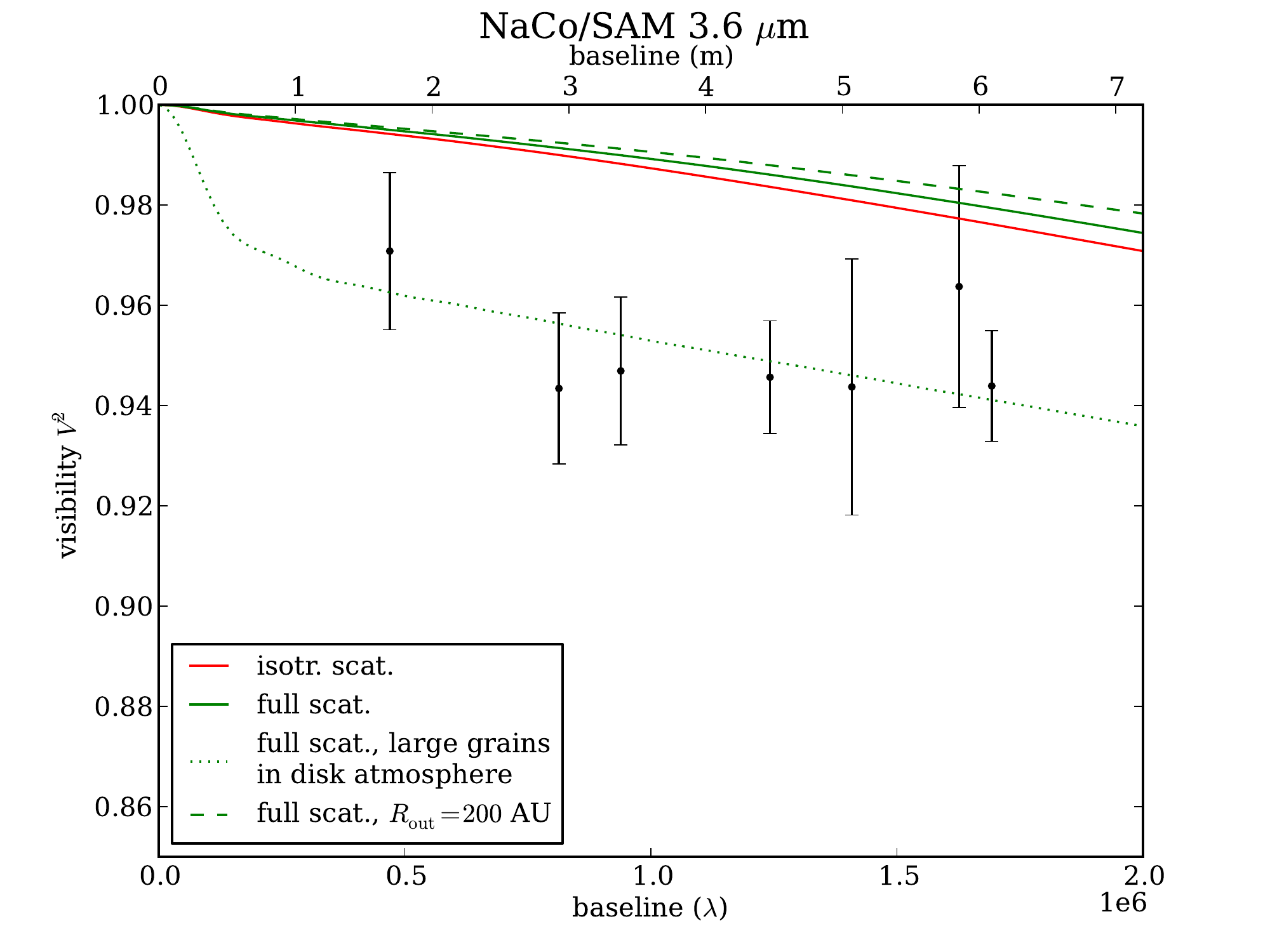}
 \caption{PIONIER and NaCo/SAM (squared) visibilities with the predicted visibilities for the final best model (\emph{full line}), under the approximation of isotropic scattering (\emph{red}) and full scattering (\emph{green}; i.e.,~also including non-isotropic scattering). The dotted green line shows the predictions for the same model, but this time with large grains in the disk atmosphere (simulated by increasing the turbulence strength $\alpha$ from $10^{-5}$ to $10^{-3}$). Finally, the dashed line again shows the predicted model visibilities of the best model, but now with an outer radius of 200\,AU for the $<10$-$\mu$m sized grains. }\label{fig:nearinfrared}
\end{figure*}

\section{Near-infrared analysis}\label{sect:nearinfrared}
For reasons given in Sect.~\ref{sect:dataselection}, we have excluded the near-infrared data from the radiative-transfer analysis up to this moment. We now investigate how these data and other scattered-light observations can be accommodated within our best model.

\subsection{VLTI/PIONIER and VLT/NaCo data}
In Fig.~\ref{fig:nearinfrared}, we plot the (squared) visibility vs.~baseline corresponding to the VLTI/PIONIER and NaCo/SAM observations. Directly calculating the visibilities at $\lambda=1.65\,\mu$m (PIONIER) and $\lambda=3.6\,\mu$m (NaCo/SAM) for the final best model in Sect.~\ref{sect:modeling} gives the result shown in red. Although the model curves qualitatively reproduce the trend in the data, the model underestimates or overestimates the global visibility level of the PIONIER or NaCo/SAM data, respectively. 

All radiative-transfer simulations made up to this moment were based on an isotropic treatment of the radiation field. Given the importance of scattered light in interpreting the near-infrared data (see Sect.~\ref{sect:scattering}), we now recalculate the same model based on full (anisotropic) radiative transfer. A correct treatment of scattering, indeed an intrinsically anisotropic process, leads to visibilities that are clearly significantly higher than the visibilities calculated under isotropic scattering (Fig.~\ref{fig:nearinfrared}). This increase in disk--star contrast\footnote{To first order, the anisotropic scattering does not change the relative scattering intensity by the different disk regions, as evidenced by the qualitatively similar visibility profiles under both scattering approximations. The change in visibility level therefore only indicates a decrease of the total disk flux with respect to the stellar flux. Mathematically, we have (assuming $V_\star\simeq1$ and $F_\mathrm{disk}/F_\star\ll1$)
\begin{eqnarray*}
 V_\mathrm{tot}=\frac{\frac{1}{F_\mathrm{disk}}V_\star+\frac{1}{F_\star}V_\mathrm{disk}}{\frac{1}{F_\mathrm{disk}}+\frac{1}{F_\star}}
 \simeq\frac{1+\frac{F_\mathrm{disk}}{F_\star}V_\mathrm{disk}}{1+\frac{F_\mathrm{disk}}{F_\star}}\simeq 1-\frac{F_\mathrm{disk}}{F_\star}(1-V_\mathrm{disk}).
\end{eqnarray*}} is a consequence of the forward scattering in combination with the pole-on orientation of the TW\,Hya disk. Indeed,  grains in the disk atmosphere will tend to scatter the stellar radiation in the radial rather than the perpendicular (i.e.,~the observer's) direction, making the disk fainter (see also \citealt{2013A&A...549A.112M}).

Taking anisotropic scattering into account improves the agreement of the model with the PIONIER data.\footnote{It is worthwhile to mention that the anisotropic scattering has no appreciable effect on the disk structure, as far as the fit to the other data is concerned.} For the NaCo/SAM data, also the full-scattering model fails to reproduce the visibilities. The latter might point to the need for grains with different scattering properties.

To further investigate this, we recalculate the visibilities for an increased amount of large grains in the disk atmosphere. Instead of adding other grains to the model, we artificially stir-up large grains by increasing the turbulent mixing strength (from $\alpha=10^{-5}$ to $10^{-3}$). As shown in the right panel of Fig.~\ref{fig:nearinfrared}, this leads to a much better reproduction of the SAM data, a consequence of the higher scatter efficiency of the larger grains (resulting in a lower star--disk contrast). Of course, a real increase in turbulence is excluded by the fit of the other data (since this completely changes the vertical distribution in all grain populations), and also the reproduction of the PIONIER data is affected. Yet, this test shows that adding a small, separate population of micron-sized grains to the upper atmosphere might allow us to fully incorporate the SAM observations. Particles like this could remain high up in the atmosphere when they are fluffy (like ``snowflakes''), i.e.,~when they have large surface-to-mass ratios and hence a strong coupling to the gas \citep{2013A&A...549A.112M}.

\subsection{Outer radius in scattered light}\label{sect:outerradius}
The outer radius of the disk ($R_\mathrm{out}$) is consistently found to be $\sim60\,$AU. This value is clearly smaller than the outer detection radius of the disk in scattered light, which is at least $280\,$AU \citep{2005ApJ...622.1171R}. However, convergence plots of $R_\mathrm{out}$ indicate that the parameter is mainly constrained by the SMA data, i.e.,~by the emission of large grains. For the small grains, which will dominate the scattering since they are dominating the disk atmosphere (due to settling), we do not have any strong constraints on $R_\mathrm{out}$ in our data. 

\citet{2012ApJ...744..162A} already indicated that the outer radius of the disk in large grains is much smaller than it appears in gas emission (as traced by CO). In the current scenario, the small grains seem thus well coupled to the gas, and span a region up to at least 200\,AU in radius, whereas the large grains are found only up to 60\,AU.  As a test, we can therefore add small grains to the region outside 60\,AU, hereby effectively introducing a decoupled $R_\mathrm{out}$ for the small grains. In principle, this should not influence the model fit to the thermal emission data.

To the plots in Fig.~\ref{fig:nearinfrared}, we added the corresponding model plots where we let the surface density of the grains smaller than $10\,\mu$m continue up to 200\,AU. The changes to the model curves are low, and clearly such a change is not detectable with our near-infrared data sets. Making corresponding images shows that, at least qualitatively, we can get synthetic near-infrared images with large outer radii, but for which the fit of the here-presented data set (near-infrared to cm interferometry) is not changed.

Recently, \citet{2013ApJ...771...45D} investigated the depression in the surface brightness of the disk at $R\sim80\,$AU, already reported in \citet{2000ApJ...538..793K}. The depression is interpreted as a (partially filled) gap in the disk. Since the detection is based on the scattered light of the disk, we lack any strong diagnostics to reconfirm this gap. Our disk model also does not require any material to be located that far from the central star, i.e.,~we currently cannot confirm if a gap would be present in the thermal dust emission.

\section{Summary and conclusions}\label{sect:conclusion}
Although the transition disk around TW\,Hya has been studied in detail, the variety of disk models in the literature indicates that the disk structure is still under discussion. Combining high-angular-resolution data that span five orders of magnitude in wavelength, our aim was to develop a radiative-transfer model that fully incorporates the dust distribution around the star, with a focus on the inner-disk region. This work has led to the following conclusions:
\begin{enumerate}
 \item Reimplementing existing radiative-transfer models for the TW\,Hya disk shows that the published models can indeed explain part of the combined data set, but fail to reproduce data at other wavelengths. This shows that a multiwavelength analysis is essential to fully characterize the disk. Also, this exercise already points to some model properties: the need for large ($\gg\mu$m) dust grains, the importance of full radiative transfer for explaining the disk, and the fact that the new mm-interferometry gives completely new constraints on the disk structure.
 \item A simple disk structure with a vertical inner rim and a radially homogeneous composition (from small to large grains) cannot properly explain the data. From the MIDI data, we infer an inner region with a lower surface density. 
 \item As a first modification to the simple disk model, we introduce a rounded rim by exponentially reducing the surface density (with respect to a standard power law surface density) inwards of a specified radius. \citet{2013A&A...557A..68M} have shown that this rim profile follows from hydrodynamical simulations of a low-mass companion opening a disk gap. Our MIDI data agree with the expected mid-infrared emission of such a rim. Although linking the gap in the TW\,Hya disk with the presence of a companion is tentative, other mechanisms might lead to similar rim geometries.
 \item The mm-emission as observed with VLA originates from a spatially more compact region than the sub-mm emission observed with SMA. We interpret this as evidence for a more compact distribution of the largest dust grains, something that is also seen in other disks (e.g.,~\citealt{2012ApJ...760L..17P}). We model the $>100$-$\mu$m grains with the same surface density profile as the smaller grains, but with a steeper surface density power. This allows us to reproduce both the SMA- and VLA-visibilities, to a reasonable extent. We note that for TW\,Hya, the VLA data allow us to trace the imprint of dust emission to wavelengths as long as 4 cm, in line with earlier suggestions by \citet{2005ApJ...626L.109W}.
 \item Our final disk model (Fig.~\ref{fig:final_drawing}) has an inner radius of $0.3-0.5\,$AU, an outer radius of $\sim60\,$AU, and a maximum in the surface density around $2.5\,$AU, where the surface density smoothly decreases inwards.
 \item For the radiative-transfer analysis, we do not focus on the near-infrared data, since these depend predominantly on the scattering properties of the dust. Instead, we compare the visibilities predicted by the best-fit model with the actual data. This comparison shows the importance of taking the full scattering properties of the dust into account. The predicted visibilities reproduce the H-band data reasonably well; for the L'-band data, a small population of possibly fluffy grains might be needed in the disk atmosphere for predicting the correct visibilities. A population of small grains at large distances from the star, as observed in scattered light images, can easily be incorporated in the final disk model. 
\end{enumerate}

\begin{acknowledgements} 

J.~Menu wishes to thank B.~Acke for useful discussions, K.~Johnston for support with the CASA software, and P.~Degroote for providing SED analysis/fitting software. We are grateful to I.~Pascucci and to the anonymous referee for comments that helped improving the manuscript. J.~Menu acknowledges an FWO travel grant for a long research stay abroad (V448412N). F.~M\'enard acknowledges support from the Millennium Science Initiative (Chilean Ministry of Economy), through grant ``Nucleus P10-022-F''. F.~M\'enard also acknowledges funding from the EU FP7-2011 under Grant Agreement No 284405. \\
The Submillimeter Array is a joint project between the Smithsonian Astrophysical Observatory and the Academia Sinica Institute of Astronomy and Astrophysics and is funded by the Smithsonian Institution and the Academia Sinica. The National Radio Astronomy Observatory is a facility of the National Science Foundation operated under cooperative agreement by Associated Universities, Inc.
 
\end{acknowledgements}

\appendix

\section{Genetic-algorithm formalism}\label{appendix:GA}
In a genetic algorithm, \emph{generations} of models are calculated starting from an initial model population with random parameters. Each subsequent generation is based on the best-fitting models of the previous generation (\emph{parents}). Gradually, the population of models is therefore evolving towards a ``fitter'' population, and the fitting process is stopped when a sufficiently well fitting individual (as defined by a criterion) is found. 

We used a formalism based on the algorithm of \citet{2009A&A...501.1259C}, with refinements as suggested by \citet{2011AJ....141...78C}. In the algorithm, the offspring generation is based on randomly picking models within the neighborhood of the fittest individual models of the parent generation. For the definition of the parameters of a child model in generation $n$, we do the following steps:
\begin{enumerate}
 \item Calculate the fitness of all parent models in generation $n-1$, where we define the fitness as
 \[\mathrm{fitness}=1/\chi^2,\]
 \item Pick a random parent model $p$, with a probability proportional to the square of its fitness, i.e.,~$(1/\chi^2)^2$,
 \item Define each child parameter $k$ as a random number from the Gaussian distribution
 \[\mathcal N\Big(k_p,\big(\sigma_{k,n-1}/2^{n/l}\big)^2\Big),\]
 where $k_p$ is the parameter value of the parent $p$, $\sigma_{k,n-1}$ the standard deviation of values of the parameter $k$ in the previous generation ($n-1$) and $l$ a fixed (positive) number.
\end{enumerate}
The choice of the above standard deviation ensures that we explore the neighborhood of interesting models in a decent way. Parameters that are easily constrained will naturally evolve quicker to their fittest value, and hence the population standard deviation $\sigma_{k,n}$ of the parameter will rapidly diminish. The factor $2^{n/l}$ will gradually speed up the confinement of the random-pick process. We fix value $l=10$, which we found appropriate for getting a sufficiently fast convergence. The number of models per generation is fixed to $200$, of which the best $10\%$ is parent for the next generation. We note that parents are also copied to the next generation, assuring that well-fitting individuals are kept in the model population (\emph{elitism}). 

\section{Error estimation}\label{appendix:error}
We start from the reasonable assumption of Gaussian errors on the data points. To each data point $(x_i,y_i)$, we can associate a Gaussian error distribution
\begin{equation}
 p(y_i|\boldsymbol\xi)\propto\exp\left(\frac{-\big(y_i-f(x_i;\boldsymbol\xi)\big)^2}{2\sigma_i^2}\right),\label{eq:prob}
\end{equation}
where $\sigma_i$ is the standard deviation of $y_i$, and $f$ is the model with parameters $\boldsymbol\xi$. The probability associated with the complete data set, i.e.,~the \emph{likelihood} function, is then
\begin{equation}
 P(\textbf y|\boldsymbol\xi)=\prod_ip(y_i|\boldsymbol\xi)\propto\exp(-\chi^2/2),\label{eq:likelihood}
\end{equation}
i.e.,~the product of the (independent) probabilities in Eq.~(\ref{eq:prob}). Here, $\chi^2$ follows the classical chi-square definition
\begin{equation}
 \chi^2=\sum_i\left(\frac{\big(y_i-f(x_i;\boldsymbol\xi)\big)^2}{\sigma_i^2}\right).
\end{equation}
In terms of our model fitting, this $\chi^2$ corresponds to our $\chi^2$ in Eq.~(\ref{eq:totchi2}). 

Knowing the likelihood function in Eq.~(\ref{eq:likelihood}), we have an estimate on how likely it is to obtain a data set $\mathbf y$ for a \emph{given} set of parameters $\boldsymbol\xi$. Bayes' Theorem allows us then to ``flip'' this relation, and find the probability distribution of our parameters \emph{given} our data set, which is what we are actually interested in:
\begin{equation}
 P(\boldsymbol\xi|\textbf y)=P(\textbf y|\boldsymbol\xi)\,P(\boldsymbol\xi)/P(\textbf y).\label{eq:bayes}
\end{equation}
Here, $P(\mathbf y)$ is the probability of the data under all possible model realizations, a factor that can simply be considered as a normalization constant (i.e.,~independent of $\boldsymbol\xi$). $P(\boldsymbol\xi)$, on the other hand, is the prior probability distribution of the parameters. In the absence of any natural preference for the parameters prior to obtaining the data, we can assume flat priors, i.e.,~uniform probability distributions for each of the parameters. In the range of possible parameter values, Eq.~(\ref{eq:bayes}) thus indicates that $P(\boldsymbol\xi|\textbf y)$ and $P(\textbf y|\boldsymbol\xi)$ are \emph{identical} under our assumptions. 

In essence, what we want is to have a sufficient amount of information on our likelihood function, or equivalently on our $\chi^2$-function (Eq.~(\ref{eq:likelihood})). Under that condition, we know the behavior of the joint probability distribution $P(\boldsymbol\xi|\textbf y)$, which implies we can calculate marginal distributions $P(\xi_i|\textbf y)$ for each of the different parameters $\xi_i$, and get expectation values and standard deviations. The latter values are then the formal error bars on each of the individual parameters.

Sampling the $\chi^2$-function (Eq.~\ref{eq:likelihood}) can be achieved by calculating models on an extended grid of parameters (e.g.,~\citealt{2008A&A...489..633P,2012A&A...539A..17L}). The grid method has the advantage that the parameter space is sampled in a regular way, making it straightforward to calculate the marginal probability distributions of the different parameters. In the case of a genetic fitting algorithm, we indeed also have a sampling of the $\chi^2$-function, though very irregular. However, the algorithm will naturally constrain the sampling to the interesting parts of the parameter space, where the sampling is also increased. For a sufficiently converged fitting algorithm, one can therefore imagine the sampling of the algorithm to be good enough for having the necessary information on the $\chi^2$-function. 

The way we estimate our modeling errors is to make a multi-dimensional interpolation of our $\exp(-\chi^2/2)$ values on a parameter grid, normalizing the sum to $1$ (in order to get a probability distribution). This is then our estimate of $P(\boldsymbol\xi|\textbf y)$, which we can marginalize to get the probability distributions for the individual parameters.

\section{Long-wavelength SED}\label{appendix:longSED}

\begin{figure}
\includegraphics[width=.5\textwidth,viewport=15 0 540 430,clip]{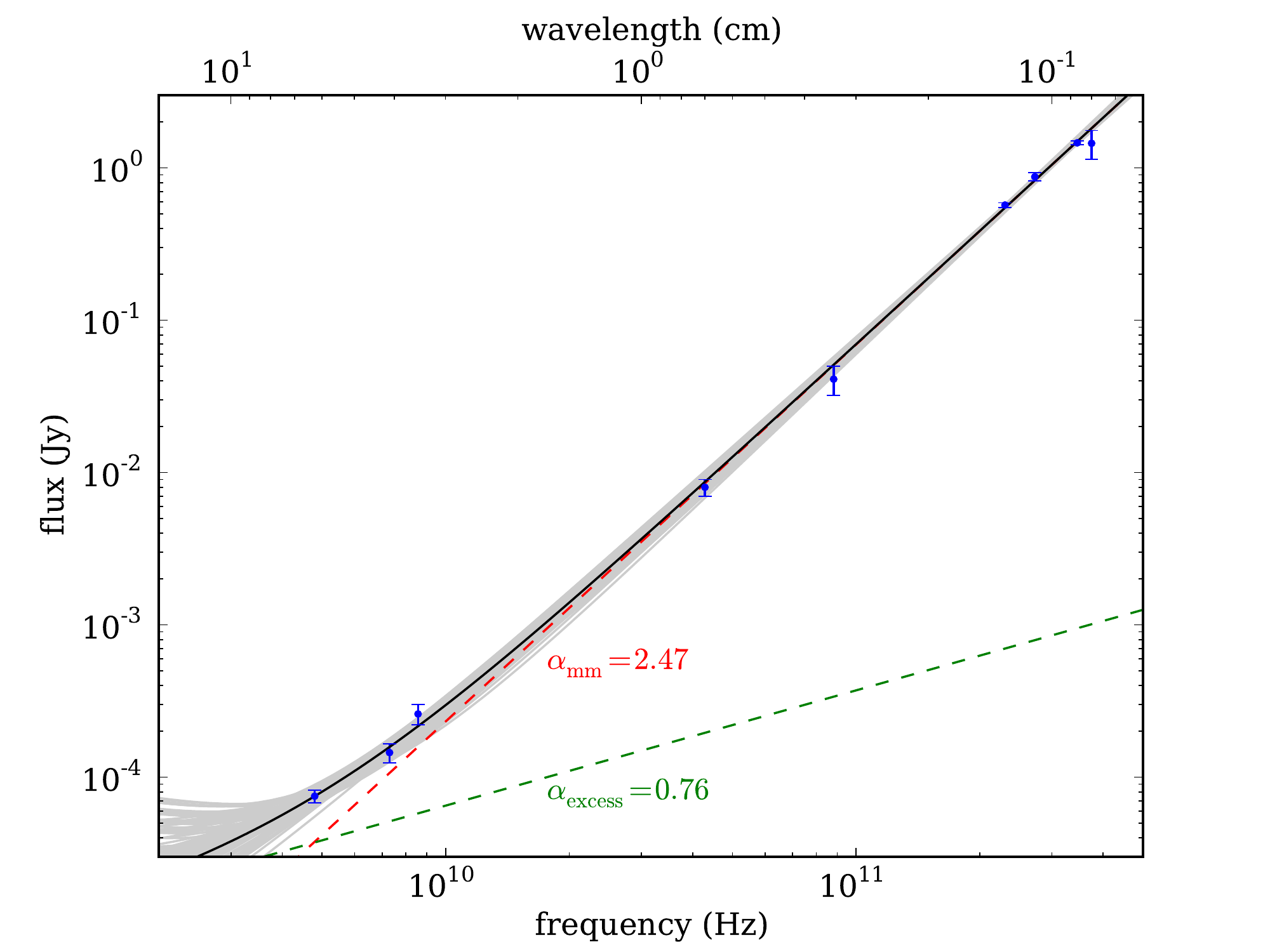}
 \caption{Long-wavelength SED of TW\,Hya, including the new $4.1$-cm and $6.3$-cm detection. The dashed lines denote the power law fits corresponding to the dust emission (\emph{red}) and the radio continuum excess (\emph{green}). The full line is the sum of the two contributions. In gray, a set of 100 models from the Monte Carlo resampling analysis is shown. }\label{fig:excess}
\end{figure}

Following the approach of \citet{2012ApJ...751L..42P}, we first fit the (sub-)mm SED with a single power law ($F_\nu\propto\nu^\alpha$), corresponding to the optically thin dust emission (Fig.~\ref{fig:excess}). This gives a spectral index of $\alpha_\mathrm{mm}=2.47\pm0.05$ (cf.~$\alpha_\mathrm{mm}=2.57\pm0.06$ found by \citet{2012ApJ...751L..42P}). As a second step, we add a second power law to the fit for estimating the excess spectral index $\alpha_\mathrm{excess}$ at cm-wavelengths. The resulting spectral slope of the best fit is $\alpha_\mathrm{excess}=0.8^{+0.4}_{-0.6}$. We applied a Monte Carlo resampling technique (see, e.g.,~\citealt{2010arXiv1009.2755A}) for inferring the uncertainty on this parameter. The inferred probability distributions for the dust-continuum and radio-excess emission allow us to estimate the contribution of the excess emission to the 9-mm VLA data. The ratio of excess to dust emission at $9.3\,$mm for the best fit is $3.8\,\%$, and it is below $12\,\%$ at a $95$-$\%$ confidence level.

On the basis of this fit to the radio excess, we thus conclude that dust emission accounts for at least $90\,\%$ (and likely even more) of the flux observed at 9\,mm. In case this emission is free-free thermal emission, the analysis of \citet{2012ApJ...751L..42P} shows that the expected radio spectral index is $\alpha_\mathrm{excess}=-0.1$. This negative spectral index suggests an even lower contribution of an excess at 9\,mm.

In their analysis of the 9-mm VLA data of the T\,Tauri star AS\,209, \citet{2012ApJ...760L..17P} find a point-like contribution to the flux on the level of $\sim10\,\%$, which they associate with free-free emission from a compact ionized wind. In our TW\,Hya data, probing significantly longer baselines, no point-source emission is detected: the averaged visibility on baselines $>1400\,$k$\lambda$ is effectively zero ($0.022\pm0.024\,$mJy). Any point-like free-free emission on the level of a few percent is thus excluded.

In a related note, we want to emphasize the interesting behavior of TW\,Hya's SED in the centimeter wavelength range and its implication. It was recognized early on that the dust emission of circumstellar disks might be traceable beyond the classical mm wavelength range ($\lambda \lesssim 3.4$\,mm), well into the centimeter regime (e.g.,~\citealt{1993Icar..106...11M}). In the more recent past, several multi-object studies detected dusty disks at wavelengths between $7$ and $16$\,mm with VLA and with ATCA (e.g.,~\citealt{2006A&A...446..211R,2009A&A...495..869L,2012MNRAS.425.3137U}). The detected signals at longer wavelengths ($\gtrsim 2$\,cm), however, were usually interpreted as arising from an ionized gas component via free-free emission. This was corroborated by the early T Tauri disk models that predicted very low flux densities from dust at a wavelength of 3.6\,cm (cf.~\citealt{1990AJ.....99..924B,1993Icar..106...11M}). Consequently, \citet{2006A&A...446..211R}, for instance, assumed that their 2.0 and 3.6\,cm VLA detections arise from free-free emission, and used the resulting spectral index to correct their 7\,mm signal for a non-dust contribution. At ATCA, often only upper limits could be established for the flux at 3 and/or 6\,cm, preventing a robust derivation of the corresponding spectral index \citep{2009A&A...495..869L,2012MNRAS.425.3137U}. 

The C-band VLA data we have used for TW\,Hya consist, as mentioned above, of two sub-bands at 4.8 and 7.3\,GHz (6.3 and 4.1\,cm). Not only can these two data points be used in the global analysis of the long-wavelength SED as shown above, each of the two sub-bands also covers a 1-GHz-wide bandpass in frequency. The multi-frequency synthesis algorithm implemented in CASA allows us to derive a spectral index along such a 1-GHz range. Hence, we can report two additional spectral indices, $1.2 \pm 0.6$ at 4.8\,GHz, and $3.6 \pm 0.4$ at 7.3\,GHz, measured around the intensity peak. The error bars are relatively large, since the total intensity signal of TW\,Hya at these frequencies is very low. One should therefore not over-interpret the actual values. Still, a qualitatively different behavior is evident. The two spectral indices are different with high confidence, much steeper at 7.3\,GHz than at 4.8\,GHz. We interpret this finding as an indication that for TW\,Hya, even at wavelengths as long as 4\,cm, the dust emission is detected and has a considerable influence on the spectral index at this frequency. At 6\,cm, the relative contribution from the dust might level out with other emission mechanisms and will eventually fade for longer wavelengths.

\bibliographystyle{aa}
\bibliography{bibliography.bib}

\end{document}